\documentclass[conference]{IEEEtran}
\usepackage{amsmath,amssymb,amsfonts}
\usepackage{algorithm}
\usepackage{algpseudocode}
\usepackage{graphicx}
\usepackage{textcomp}
\usepackage{xcolor}
\usepackage{subcaption}
\usepackage{cite}
\usepackage{hyperref}
\usepackage[nameinlink,capitalize]{cleveref} 
\usepackage{float}
\usepackage{braket}
\usepackage{url} 
\usepackage{booktabs}

\begin{document}

\title{Generalised Circuit Partitioning for Distributed Quantum Computing}

\author{\IEEEauthorblockN{Felix Burt}
\IEEEauthorblockA{Department of Electrical and\\Electronic Engineering\\
Imperial College London\\
London, UK\\
f.burt23@imperial.ac.uk}
\and
\IEEEauthorblockN{Kuan-Cheng Chen}
\IEEEauthorblockA{Department of Electrical and \\ Electronic Engineering\\
Imperial College London \\
London, UK\\
kuan-cheng.chen17@imperial.ac.uk}
\and
\IEEEauthorblockN{Kin K. Leung}
\IEEEauthorblockA{Department of Electrical and \\ Electronic Engineering\\
Imperial College London \\
London, UK\\
kin.leung@imperial.ac.uk}}
\maketitle

\begin{abstract}
Distributed quantum computing (DQC) is a new paradigm aimed at scaling up quantum computing via the interconnection of smaller quantum processing units (QPUs). Shared entanglement allows teleportation of both states and gates between QPUs. This leads to an attractive horizontal scaling of quantum processing power, which comes at the expense of the additional time and noise introduced by entanglement sharing protocols. Consequently, methods for partitioning quantum circuits across multiple QPUs should aim to minimise the amount of entanglement-based communication required between distributed QPUs. Existing protocols tend to focus primarily on optimising entanglement costs for gate teleportation or state teleportation to cover operations between QPUs, rather than both at the same time. The most general form of the problem should treat gate and state teleportation on the same footing, allowing minimal cost circuit partitions through a combination of the two. This work introduces a graph-based formulation which allows joint optimisation of gate and state teleportation cost, including extensions of gate teleportation which group gates together for distribution using common resources. The formulation permits low e-bit cost for a variety of circuit types. Using a basic genetic algorithm, improved performance over state-of-the-art methods is obtained in terms of both average e-bit cost and time scaling.
\end{abstract}


\section{Introduction}
Distributed quantum computing (DQC) is a new paradigm aimed at connecting multiple quantum processing units (QPUs) to form larger, more powerful, quantum computing systems \cite{cuomo_towards_2020}. This provides an alternative route to scaling qubit numbers \cite{caleffi_distributed_2022} which avoids the challenges of connecting many qubits on a single quantum processor \cite{unruh_maintaining_1995}. However, DQC creates new difficulties relating to the interaction of QPUs in both hardware and software. Quantum information can be distributed between QPUs using entangled pairs of qubits, or \textit{e-bits}, which must be efficiently distributed between devices \cite{caleffi_distributed_2022}. E-bits can be consumed to teleport quantum states between QPUs, or to directly perform certain quantum gates non-locally \cite{eisert_optimal_2000}. However, preparation of e-bits is estimated to be over 10 times slower than local two-qubit gates \cite{isailovic_interconnection_2006}, resulting in a new objective for quantum circuit optimisation \textendash~minimisation of e-bit consumption. This has been referred to as the ``DQC problem" \cite{andres-martinez_distributing_2023} or the ``circuit partitioning problem" \cite{barral_review_2024}. Broadly, the aim of the problem is to cover all gates in a circuit split between multiple QPUs using as few e-bits as possible. Dominant methods in the literature for solving circuit partitioning focus on optimising e-bit cost from either state teleportation \cite{baker_time-sliced_2020, sundaram_distributing_2023, nikahd_automated_2021} or gate teleportation \cite{andres-martinez_automated_2019, andres-martinez_distributing_2023, wu_entanglement-efficient_2023, cuomo_optimized_2023, sundaram_efficient_2021}, failing to fully leverage the possibilities available to cover inter-QPU operations. While there are works which consider both \cite{ferrari_compiler_2021, ferrari_modular_2023, sundaram_distributing_2023}, they fail to treat them jointly as target of optimisation, resulting in higher e-bit costs than what is possible in general. This work proposes a generalised graph formulation of the circuit partitioning problem, aiming to unify the dominant approaches. The results are compared with leading methods from different paradigms \cite{baker_time-sliced_2020, andres-martinez_distributing_2023}, which both perform very well on circuits which fit their methods but perform sub-optimally in certain cases. Using simple heuristics, we obtain competitive performance with the best cases of existing methods, while retaining strong performance in instances where the benchmark methods break down, leading to lower e-bit costs on average. The results are obtained using a genetic algorithm with time complexity linear in the total circuit size (width $\times$ depth), with experimental demonstrations showing significant improvement over the scaling of the benchmark methods.
\section{The Circuit Partitioning Problem (CPP): Background}
\subsection{Static circuit partitioning}
The general aim of circuit partitioning is to split a quantum circuit $C$, consisting of list of operations on a set of logical qubits $q = \{q_{0},q_{1}...\}$, among a set of QPUs $Q = \{Q_{0}, Q_{1}...\}$. Splitting quantum circuits into separate parts can be considered as assignment $\phi$ of logical qubits to QPUs, $\phi : q \rightarrow Q$, resulting in a collection of operations which are either \textit{local} or \textit{non-local}. Local operations either act only on a single qubit, or act on multiple qubits which are assigned to the same QPU, i.e. $\phi(q_{i}) = \phi(q_{j})$. Non-local operations, on the other hand, are gates which involve two or more qubits assigned to different QPUs, $\phi(q_{i}) \neq \phi(q_{j})$. Any assignment $\phi$ will give a corresponding number of non-local operations to be covered. In order to minimise the number of non-local operations, a quantum circuit can be converted to an interaction graph where each qubit corresponds to a node on the graph, while interactions of all pairs of qubits are summed into weighted edges. This reduces the problem to min-cut graph partitioning \cite{ferrari_modular_2023}. For balanced partitions min-cut is NP-hard, but good heuristics are available \cite{karypis_fast_1998}. In terms of circuit partitioning, this is referred to as \textit{static} partitioning because a single qubit assignment is used for the full circuit. Static partitioning methods minimise the number of non-local operations, which is only equal to the e-bit cost if each non-local operation takes a single e-bit. In general, this is not the case, as e-bits can be used in more flexible ways.
 \begin{figure}[ht!]
    \centering
    \begin{subfigure}{0.4\textwidth}
        \centering
        \includegraphics[width=\linewidth]{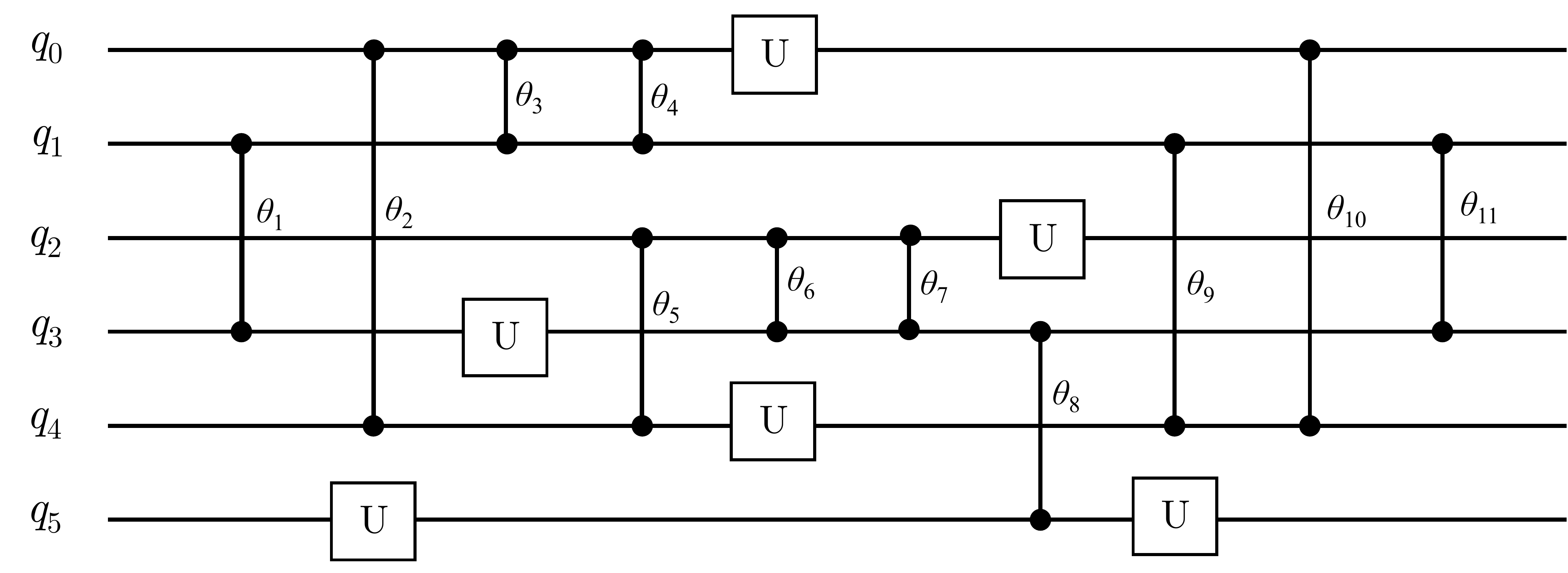}
        \label{fig:circ_static} 
        \caption{Random quantum circuit. See section \ref{sec:cp_frac}.}
    \end{subfigure}
    \begin{subfigure}{0.15\textwidth}
        \centering
        \includegraphics[width=\linewidth]{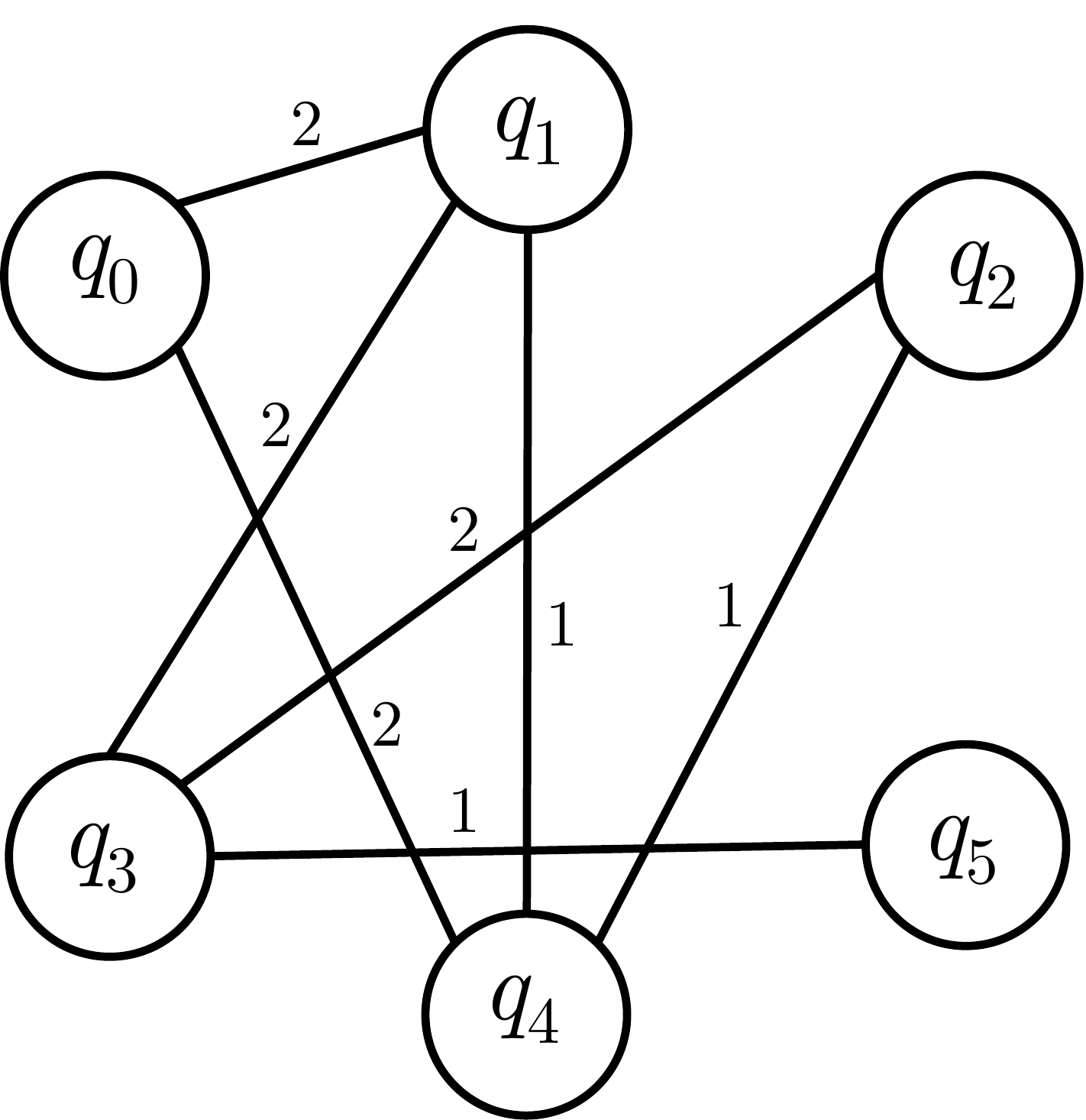}
        \label{fig:static_nopart}
        \caption{}
    \end{subfigure}
    ~~~~
        \begin{subfigure}{0.2\textwidth}
        \centering
        \includegraphics[width=\linewidth]{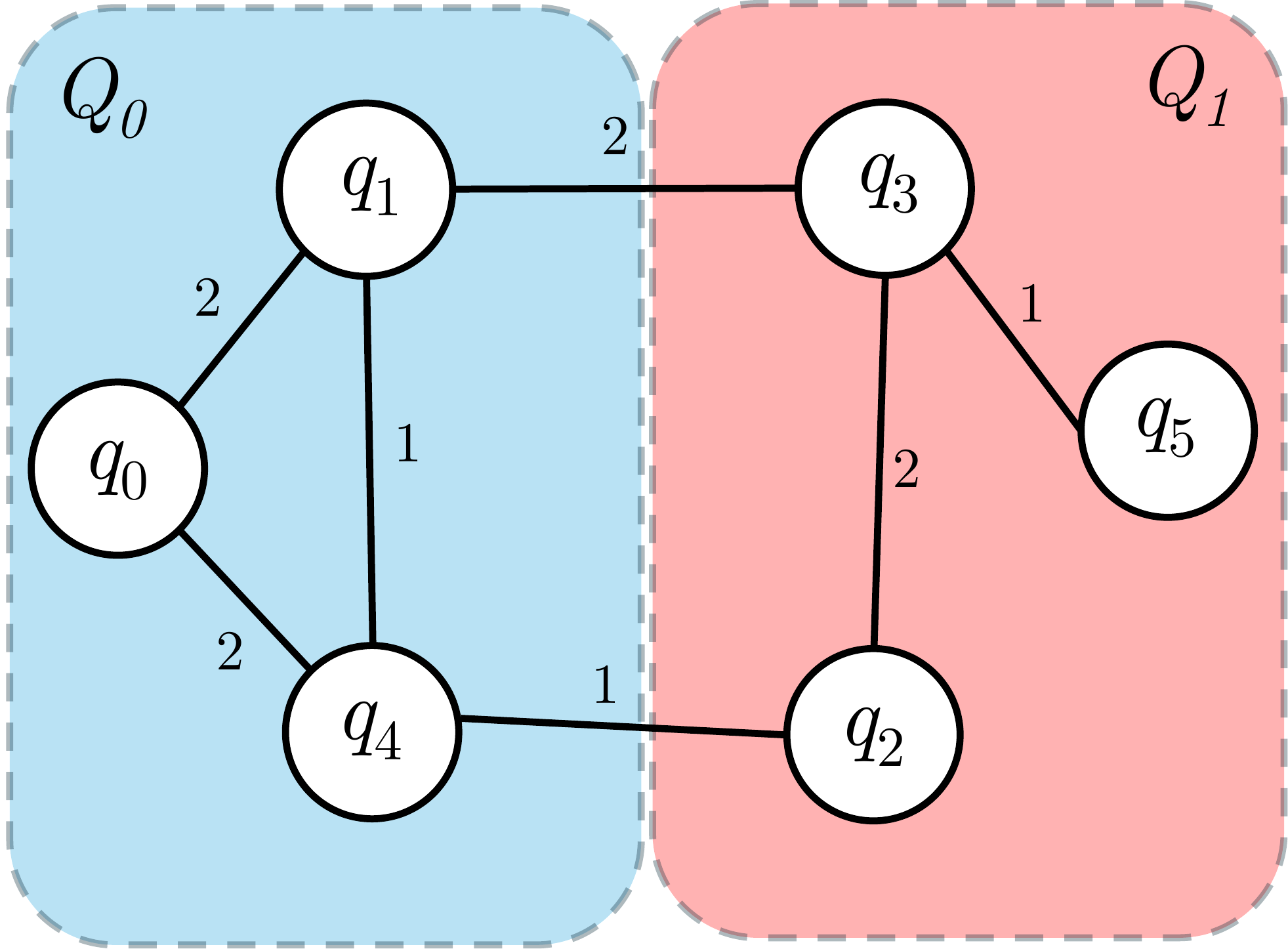}
        \label{fig:static_opt}
        \caption{}
    \end{subfigure}
    
    \caption{\textbf{Static partitioning workflow.} The circuit in a) is converted to a static interaction graph shown in b). The graph is then partitioned according to the constraints of the QPUs, such that the edge weight between partitions is minimised.}
    \label{fig:workflow}
\end{figure}

\subsection{Teleportation}
\subsubsection{State teleportation}\label{sec:state_telep}
Quantum state teleportation is the ``conventional" teleportation procedure, which uses a single e-bit, shared between a two parties, to move an arbitrary quantum state from one party to the other \cite{bennett_teleporting_1993}. 
\begin{figure}[H]
\centering
\includegraphics[width=0.85\linewidth]{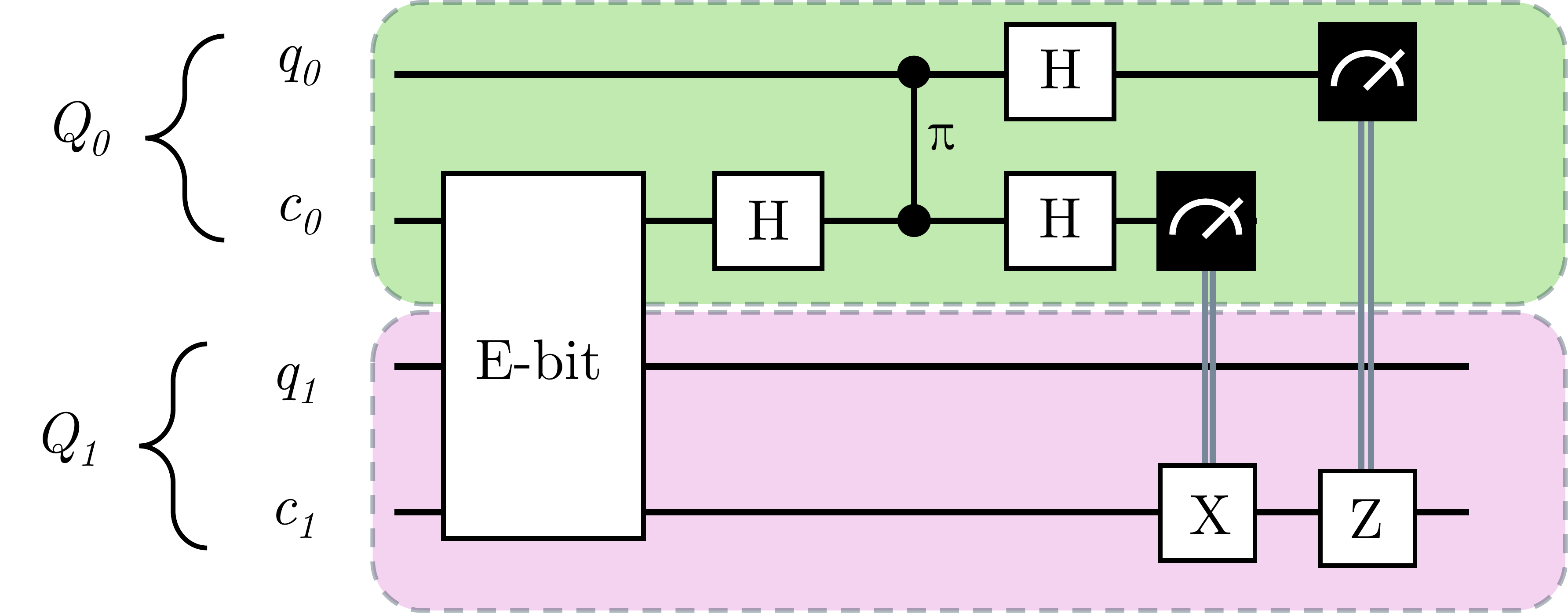}
\caption{\textbf{State teleportation.} This is the well known quantum teleportation procedure implemented using a controlled-phase gate rather than a CNOT.}
\label{fig:st}
\end{figure}State teleportation can be used to cover a non-local operation by moving one qubit state to its partner's QPU, then performing the operation locally, at the cost of a single e-bit. If the state is then returned to its original QPU, the overall e-bit cost is two. However, it is not always required to return the qubit to its original QPU \textendash ~in some cases it may even be beneficial for the qubit to remain at the destination of the teleportation. 

\subsubsection{Gate teleportation}
A less well-known procedure is gate teleportation, which uses a single e-bit to perform a two-qubit controlled unitary, without directly transferring either qubit to the other QPU \cite{eisert_optimal_2000}.
\begin{figure}[H]
\centering
\includegraphics[width=0.9\linewidth]{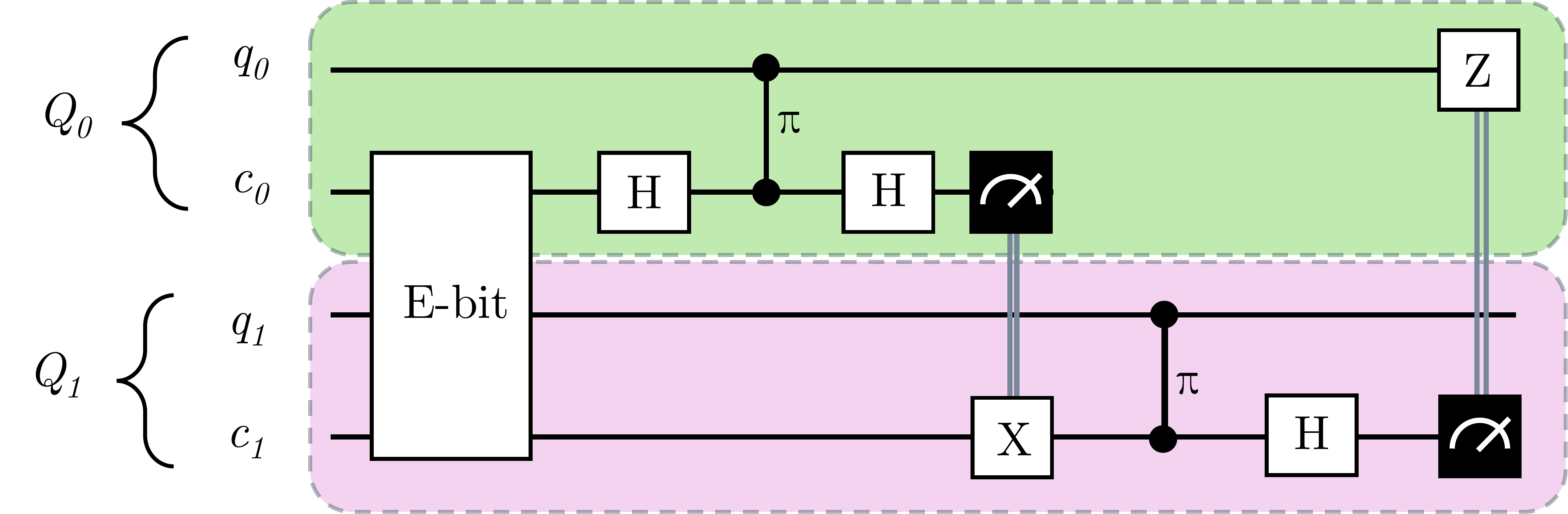}
\caption{\textbf{Gate teleportation.} The circuit displays how a single controlled-phase gate can be teleported using an e-bit. The procedure is general to all controlled-unitary gates with $q_{0}$ as the control.}
\label{fig:gt}
\end{figure}
Gate teleportation works by entangling the control qubit with a communication qubit in another QPU, such that both qubits behave the same way when controlling two-qubit gates on other qubits. The state of the control qubit is mapped from $\ket{\psi} = \alpha \ket{0} + \beta \ket{1}$ to the joint state $\ket{\psi} = \alpha \ket{00} + \beta \ket{11}$ with the communication qubit. This allows the communication qubit to control a two-qubit (or multi-qubit gate) in place of the original qubit. After the gate has been performed, the communication qubit is measured, and the result used to correct the state of the original control qubit. In contrast with state teleportation, at the end of the procedure both qubits remain assigned their original QPUs.

\subsubsection{Extended gate teleportation} \label{sec:egt}
While the basic gate teleportation procedure focuses on teleporting a single qubit controlled-unitary, it has been shown that larger sequences of operations can in fact be teleported, sometimes at no additional cost \cite{eisert_optimal_2000}. 
\begin{figure}[H]
\centering
\includegraphics[width=0.85\linewidth]{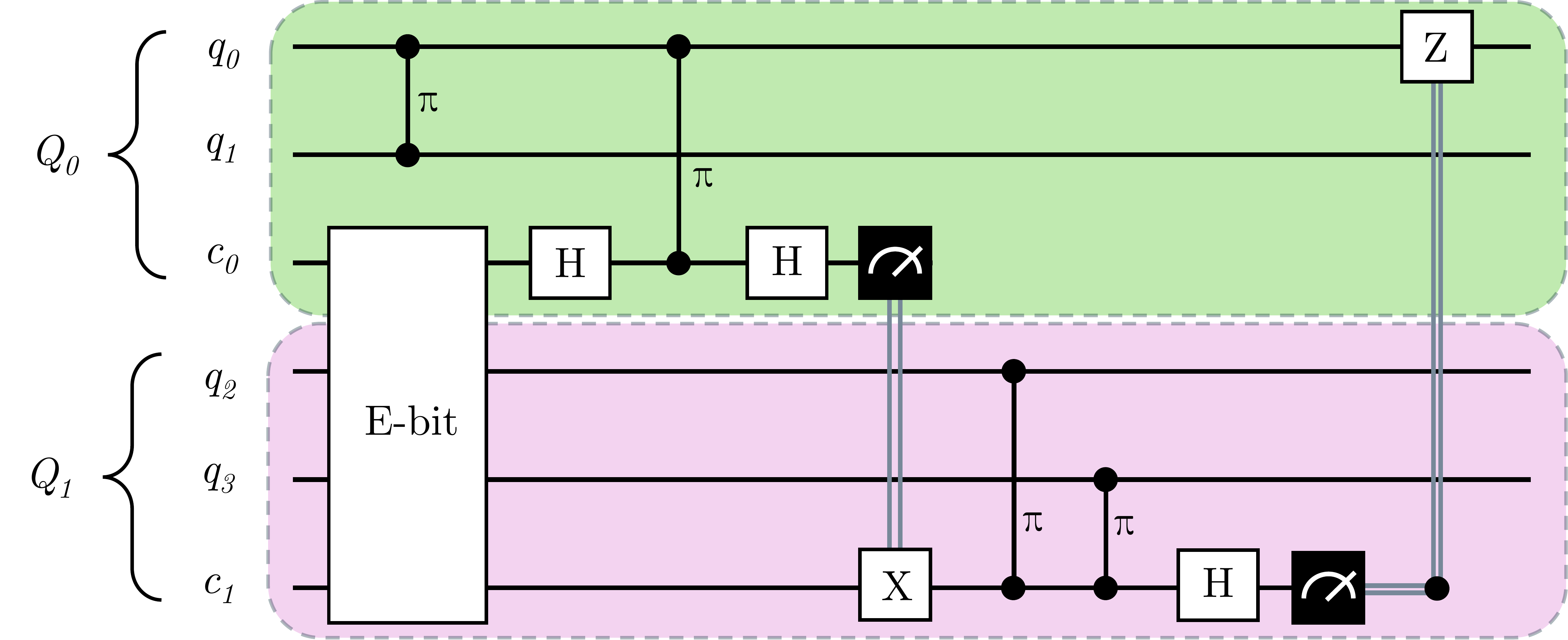}
\caption{\textbf{Extended gate teleportation.} The circuit displays how two controlled-unitary gates can be teleported using the same e-bit.}
\label{fig:egt}
\end{figure}
In particular, for a collection of two-qubit controlled-unitaries, if each gate shares a common control qubit and their target qubits are assigned to the same QPU, then all gates can be teleported using a single e-bit. This requires a delay in the measurement of the communication qubit in the target QPU, waiting until all gates in the group being covered have been completed to end the process and send the classical information back to the original QPU. In addition, any diagonal or anti-diagonal single-qubit gate acting on the common control qubit can be included in the teleportation procedure \cite{huelga_quantum_2001}, as this does not change the behaviour of the linked communication qubit as a control. 

Extended gate teleportation can be broken up into a ``starting process",  which creates the linked entangled state, an ``intermediate process" where the corresponding local operations are performed, and an ``ending process" which ends the link with measurement and classical communication \cite{wu_entanglement-efficient_2023}. The starting and ending processes have also been referred to as the``cat-entangler" and ``cat-disentangler", respectively \cite{yimsiriwattana_generalized_2004}. The groups of gates covered is referred to as a ``distributable packet" \cite{wu_entanglement-efficient_2023}. Wu et al. go on to describe how non-adjacent distributable packets can be merged together via ``embedding" distribution processes for intermediate unitaries within a larger process. Embedding processes allow further reduction of e-bit costs but are beyond the scope of the paper \textendash ~details can be found in Wu et al. \cite{wu_entanglement-efficient_2023}. Following the literature, we will also refer to extended gate teleportation methods as ``gate packing" methods. 

\subsection{Competing methods}
The dominant methods in the DQC literature tend to focus heavily on exploiting the benefits of either state teleportation, or gate packing to produce low e-bit costs. We will roughly outline these two extreme ends of the spectrum giving reference to the relevant existing works. 
\subsubsection{Fine-grained partitioning}
As an alternative to basic graph partitioning, Baker et al. \cite{baker_time-sliced_2020} propose a method referred to as \textit{fine-grained partitioning} (FGP). Rather than reduce a quantum circuit to a single, static, interaction graph, the essence of FGP is to produce a sequence of interaction graphs for each layer, or ``time-slice", of the circuit. Time-sliced graphs can consider future interactions in the circuit by adding time-decaying weighted edges for gates in later slices. Edges representing gates in the current slice are given infinite weights to ensure a local assignment at the time of interaction. The strategy is then to find a low-cost sequence of non-local swaps (two-way gate teleportation) to transition between each slice of the circuit. The result is that all non-local operations are covered by state teleportation, while the look-ahead weights work to ensure that swaps also consider future interactions in the circuit. The methods produce low cost assignment sequences in certain scenarios but are limited by the restriction that all non-local operations are covered using state teleportation, especially for circuits for which qubit connectivity changes erratically. Similar methods are presented in Sundaram et al. \cite{sundaram_distributing_2023} and Nikhad et al. \cite{nikahd_automated_2021}. The significance of FGP type methods is that the solution for the partitioning problem is no longer a single state assignment of qubits, rather a different assignment for each layer of the circuit. Partitioning over time captures gradual changes in interactivity of quantum circuits very well. 
\subsubsection{Gate packing}
A number of other methods are motivated rather by extended gate teleportation, focusing on packing sequences of gates together into the same teleportation protocol to reduce e-bit cost. A hypergraph partitioning method was proposed originally by Andres-Martinez et al. \cite{andres-martinez_automated_2019}. This work extends the normal graph approach by using hyperedges to represent groups of gates which can be distributed using a single e-bit, such that the number of hyperedges which cut the partition directly corresponds to e-bit count rather than non-local gate count. The methods are extended in Andres-Martinez et al. \cite{andres-martinez_distributing_2023} and combined with the methods from Wu et al. \cite{wu_entanglement-efficient_2023}, which extend the scope of gate teleportation further. The methods go beyond just gate packing, using the embedding methods mentioned in Section \ref{sec:egt}. The methods presented in the latter work are capable of finding very low cost distributions in instances of circuits which contain large sequences of gates meeting the conditions for gate packing and embedding, but are unable to take advantage of state teleportation when such sequences are not available, such as in the quantum volume benchmarks used in their work.

\subsubsection{Hybrid methods}
We acknowledge the existence of methods that consider both extended gate teleportation and state teleportation \cite{ferrari_compiler_2021, ferrari_modular_2023,sundaram_distributing_2023}, but note that they treat state teleportation and gate teleportation separately in the optimisation process. For example, Sundaram et al. first optimise extended gate teleportation cost using static graph partitioning, then see if the result can be improved by strategically inserting teleportations into circuits \cite{sundaram_distributing_2023} . Similarly, Ferrari et al. begin with static graph partitioning using the METIS library \cite{karypis_fast_1998} followed by a scheduling pass which chooses whether to use state or gate teleportation to cover the gates based on cost \cite{ferrari_modular_2023}. In addition, an appendix in Andres-Martinez et al. describes how large circuits are broken up into sections which are partitioned separately, requiring teleportations to transition between each section, making this method technically a hybrid method \cite{andres-martinez_automated_2019}. However, this is dealt with using a pre-processing routine to determine where circuits should be split, rather than combined treatment of state and gate teleportation.

\section{Generalised circuit partitioning (GCP)}
To capture full generality, the problem formulation should cover instances where circuits are best executed using state teleportation only, gate teleportation only, and a combination of both \textendash ~including gate packing. Doing so results in a unification between fine-grained partitioning methods and gate packing methods, thus permitting low e-bit cost in all cases. We will present two variants of GCP; a \textit{simple} variant which considers only regular, non-extended, gate teleportation, and an \textit{extended} variant which includes simple gate packing procedures.

\begin{figure}[ht!]
    \centering
        \includegraphics[width=0.4\textwidth]{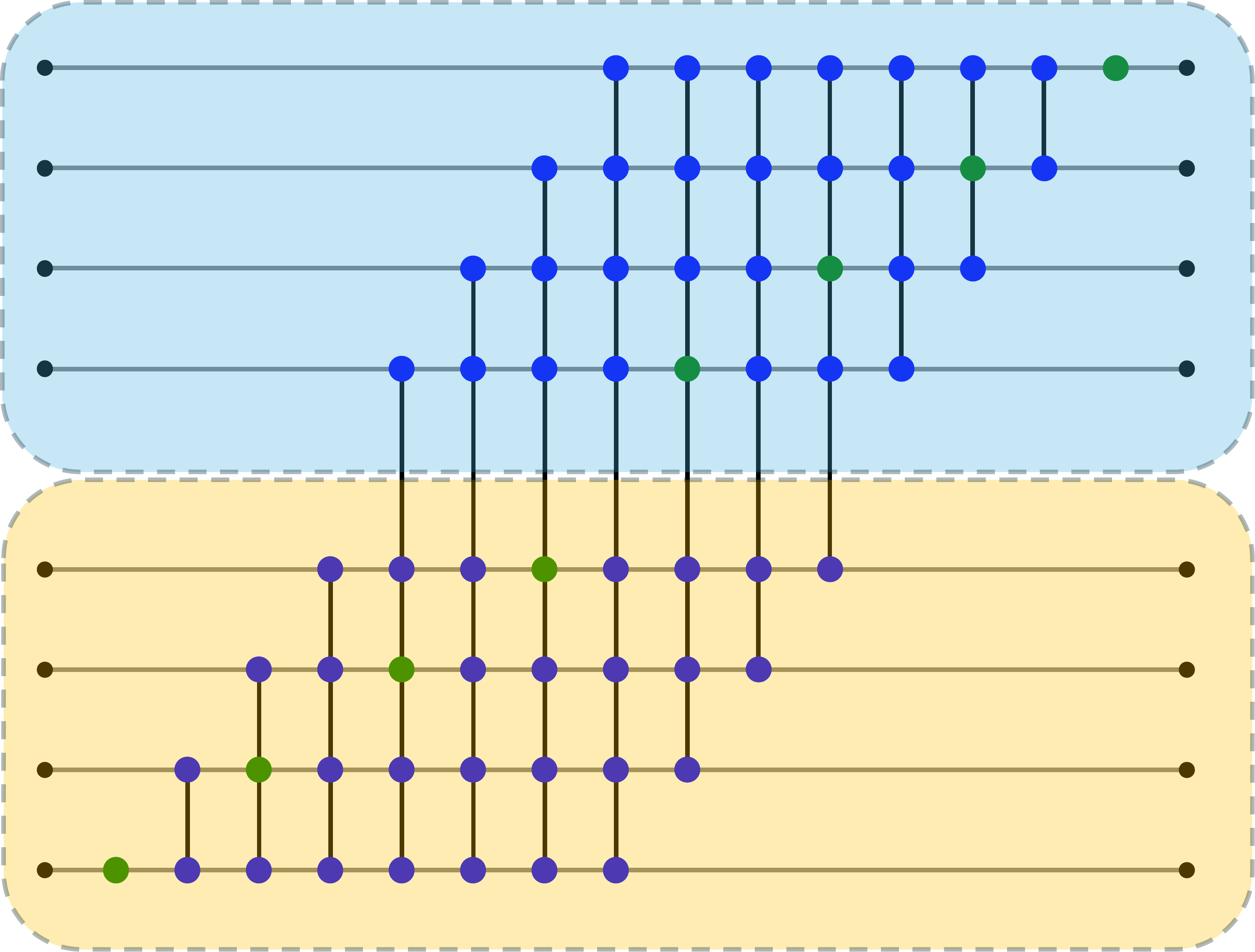}
        \caption{\textbf{Extended circuit interaction graph.} The graph shown here represented a QFT circuit on 8 qubits. Here the nodes are all assigned an unoptimised static mapping indicating gate teleportation for all edges cutting the partition.}
        \label{fig:cig_no}
\end{figure}

\subsection{Generalised circuit partitioning - simple (GCP-S)} 

Central to both methods is an extended circuit interaction graph (see Figure \ref{fig:cig_no}, which retains the time-step/layer number for all gates in the circuit. In contrast to FGP methods, which produce a different graph for each layer of the circuit, the extended interaction graph is a single graph $G(V,E)$, containing separate nodes for each qubit at each layer of the circuit. Gate-like edges connect partner nodes at the layer of their interaction, while additional state-like edges connect adjacent nodes representing the same qubit. The result is that an assignment of nodes to QPUs corresponds to an assignment of qubits for each layer of the circuit, thus determining both the state teleportations and the gate teleportations required distribute the circuit. Formally, for each logical qubit $i$ at the $l^{th}$ layer of the circuit, $q_{i}^{(l)}$ maps uniquely to a node $v_{k}$. However, valid solutions are constrained to those which always meet the QPU size constraints. These are enforced by including nodes to represent any idle qubits in each QPU, thus we increase the size of the set of logical qubits to match the total number of physical qubits. This allows the space of all valid solutions to be explored via permutations of nodes within their own layer. This picture extends the problem to a more general, constrained, graph partitioning problem. The aim is still to find a min-cut partition of the nodes but the number of nodes has been extended to reflect the dynamics of the circuit. This picture aligns closely with that presented in Davis et al. under the name ``gate partitioning" \cite{davis_towards_2023}. However, the authors do not include the constraints to enforce the size constraints of the QPUs, rather they perform post-processing to ensure that these conditions are met, resulting in high e-bit costs in many of the presented benchmarks. The authors also consider no extension to include gate packing, as is done in Section \ref{sec:gcp_e}.

\begin{figure}[ht!]
    \centering
        \includegraphics[width=0.4\textwidth]{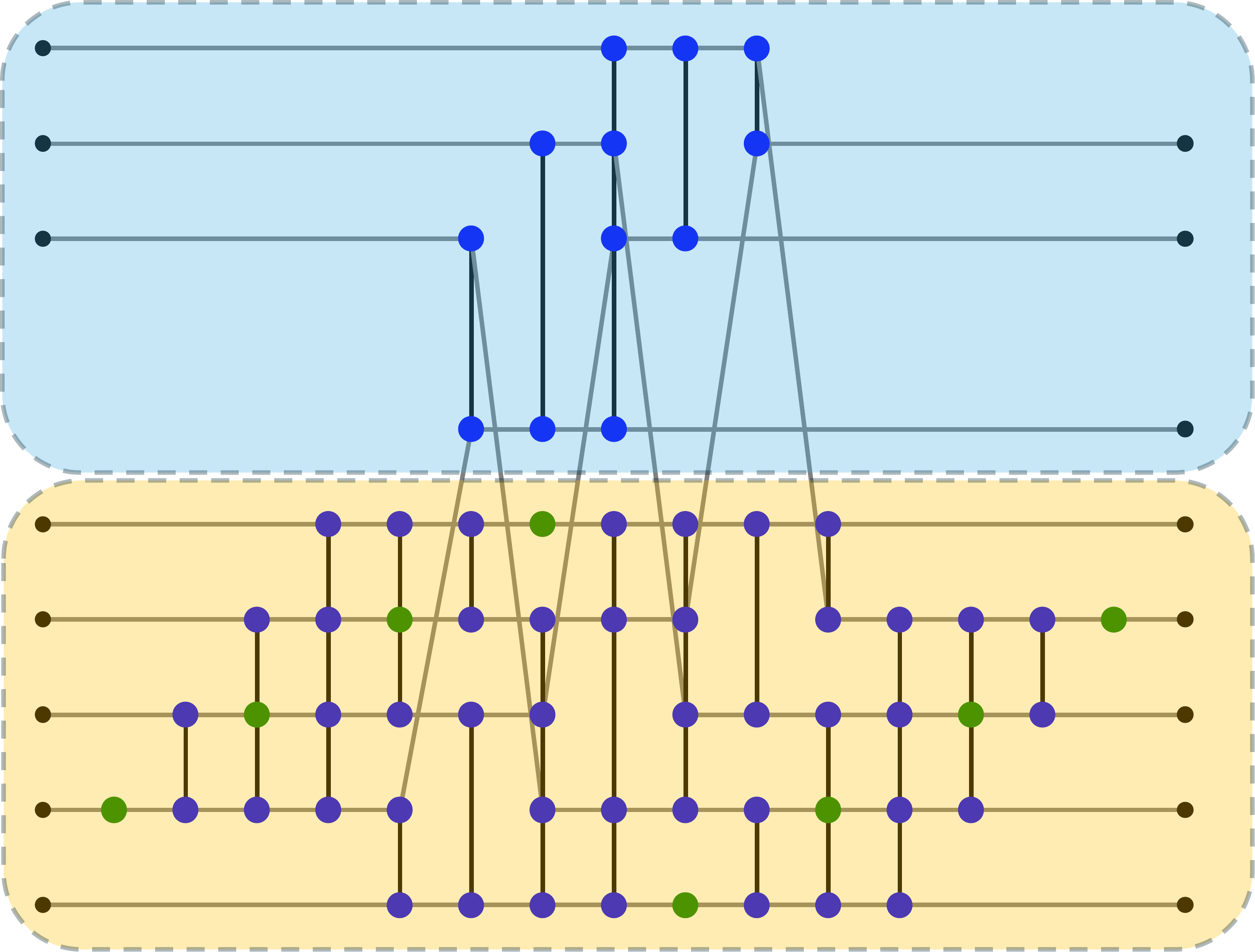}
        \caption{\textbf{Simple optimisation.} This is the same QFT as Figure \ref{fig:cig_no}, optimised according to GCP-S. Nodes have been dragged across to indicate where state teleportation occurs. The circuit is covered using 6 state teleportations, provided there is a single additional space in the second QPU.}
        \label{fig:cig_so}
\end{figure}

Formally, we define a partition of the circuit as a mapping from the set of nodes to the set of QPUs, $\Phi: V \rightarrow Q$, though since the nodes represent qubits at each layer of the circuit, we can equivalently think of a partition as an assignment of qubits to QPUs for each layer of the circuit. We retain similar notation to the static case, except that the capital $\Phi$ is used to denote assignment across all layers. As such, a partition can be viewed as a matrix where the element $\phi_{i,j}$ indicates the QPU that qubit $q_{j}$ is assigned to at layer $i$. The cost of a partition is given by:

\begin{equation} \label{cost_s}
    c = \sum_{u,v}e_{u,v}(1 - \delta_{\Phi(u)\Phi(v)})f(r(\Phi(u),\Phi(v))),
\end{equation}
where we have included a generic function $f$ to denote the cost scaling, as a function of the length $r(\Phi(u),\Phi(v))$ of the path separating the nodes $u$ and $v$ in the network. We only consider homogeneous networks, for which $f \rightarrow 1$ in all cases, though it is included to demonstrate ease of extension.

 \begin{figure}[ht!]
    \centering
    \begin{subfigure}{0.24\textwidth}
        \centering
        \includegraphics[width=\linewidth]{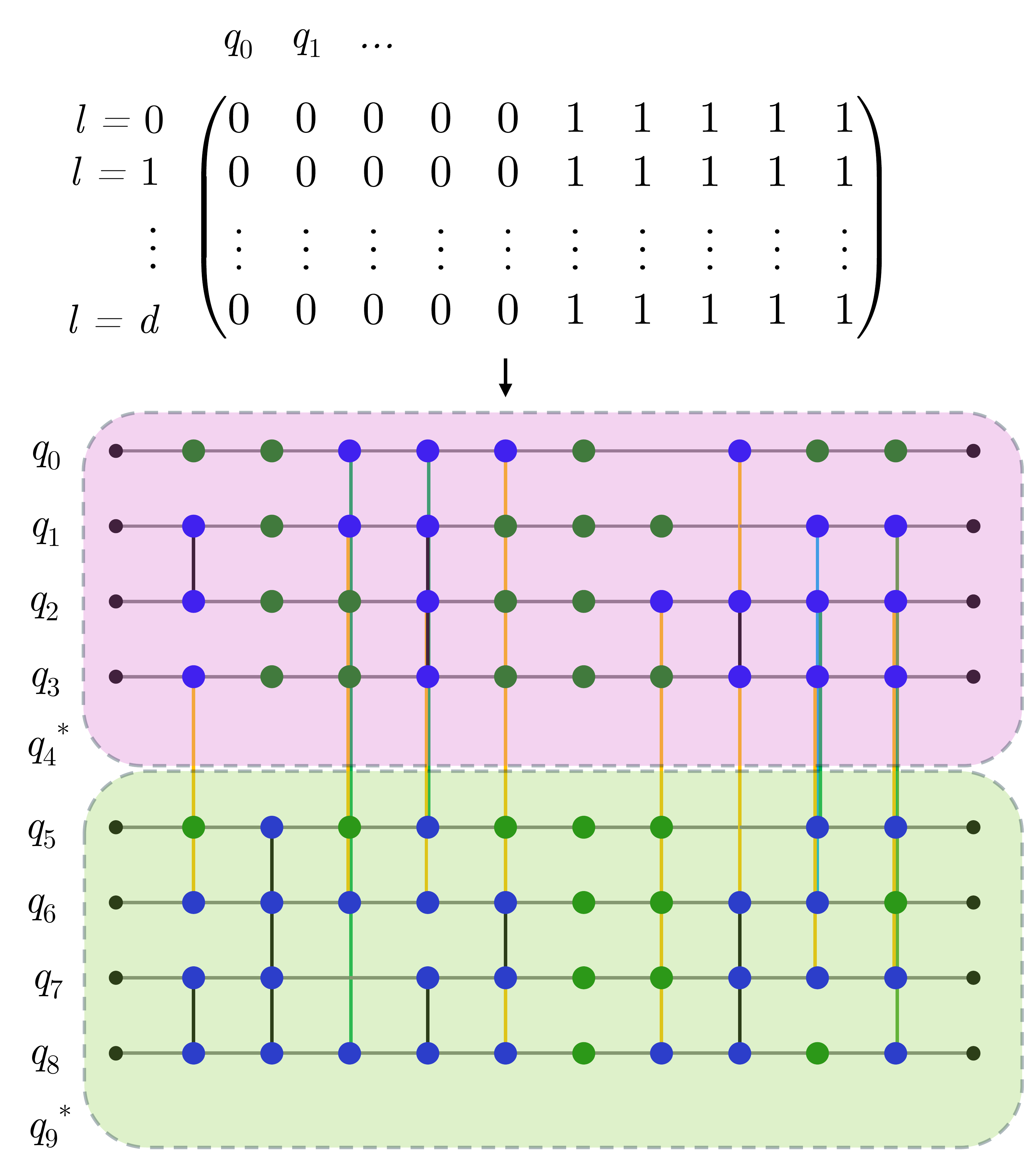}
        \label{fig:partition_no_opt} 
    \end{subfigure}
    \begin{subfigure}{0.24\textwidth}
        \centering
        \includegraphics[width=\linewidth]{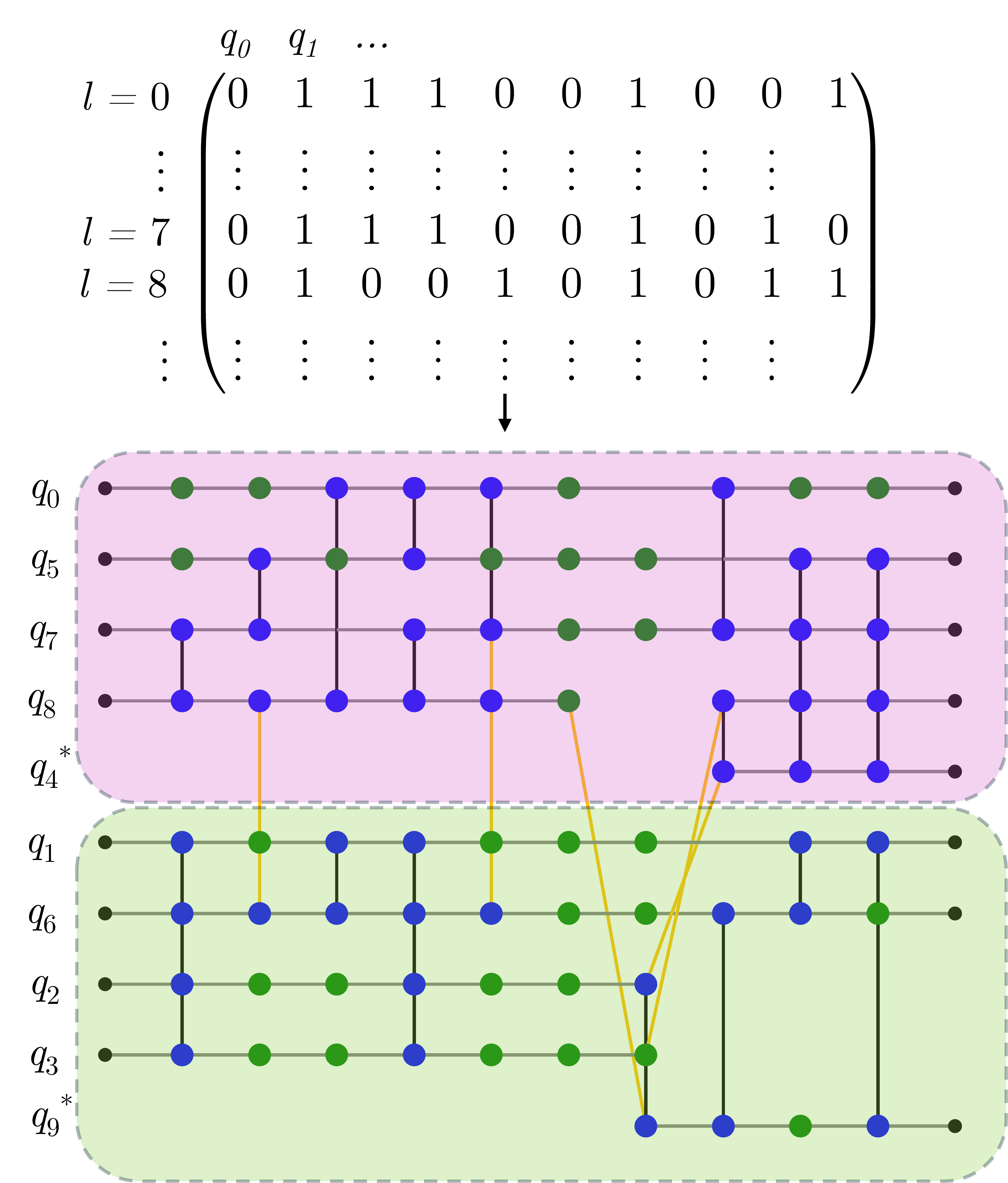}
        \label{fig:partition_opt}
    \end{subfigure}
    
    \caption{\textbf{Visualising assignments on graphs.} In a) we see the cost of a naive assignment, resulting in an e-bit cost of 13. In b) we see the effect of an optimised assignment sequence, resulting in an e-bit cost of 5. Coloured nodes indicate where e-bits are used.}
    \label{fig:assignment}
\end{figure}

\subsection{Generalised circuit partitioning - extended (GCP-E)}\label{sec:gcp_e}

 The extension to GCP can be achieved by modifying the edge structure of the graph. The graph is first scanned to identify where gates can be grouped together according to the conditions identified in Section \ref{sec:egt}. Since different possibilities are available for grouping gates, a greedy approach is used, which always adds gates to the largest available group. The conditions for grouping gates are:
 \begin{enumerate}
     \item \textbf{All two-qubit gates share a common control qubit.}
     \item  \textbf{All two-gates are adjacent on the control.} 
 \end{enumerate}

 \begin{figure}[ht!]
    \centering
        \includegraphics[width=0.4\textwidth]{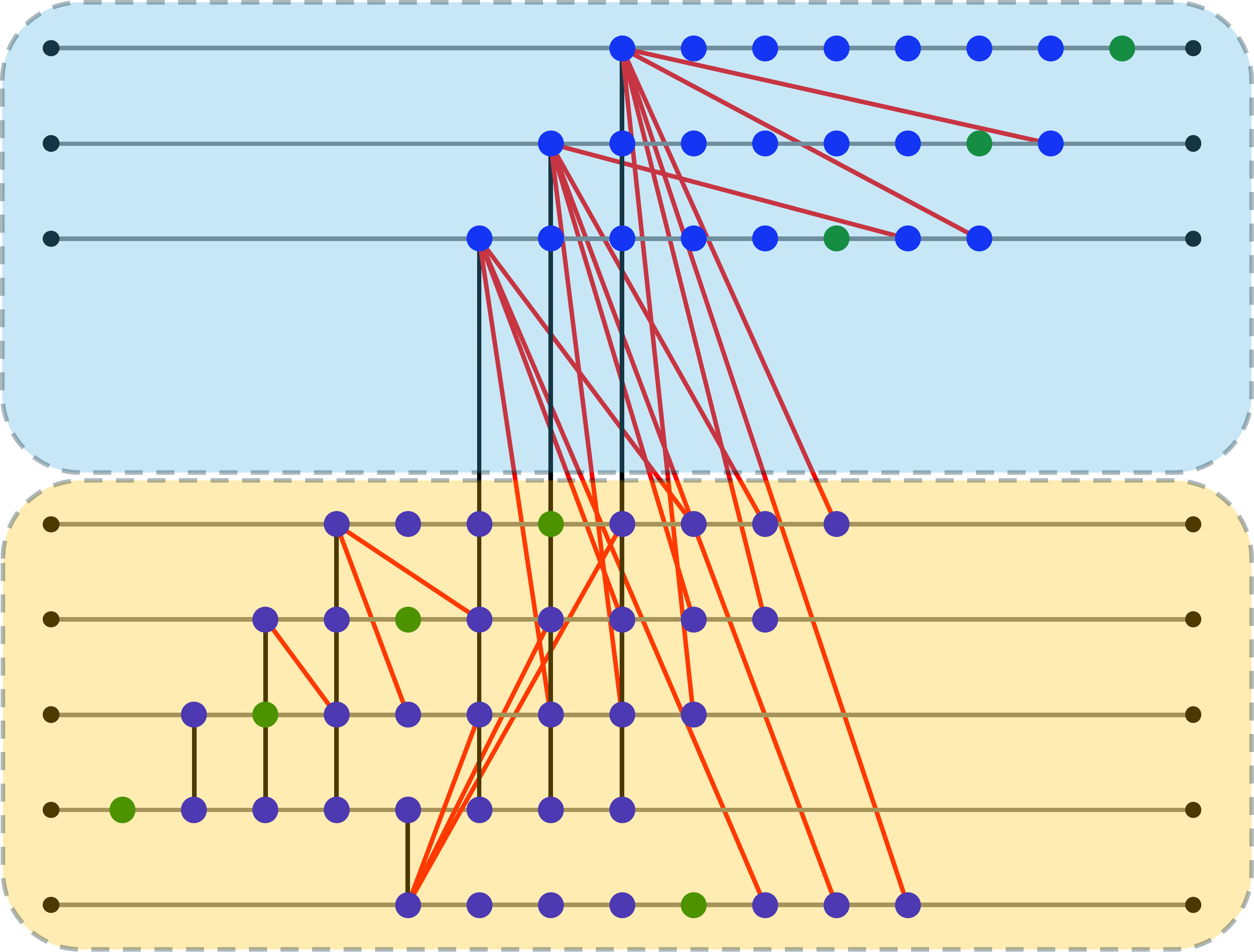}
        \caption{\textbf{Extended optimisation.} The Quantum Fourier Transform (QFT) circuit optimised using the GCP-E formulation. In this case, no state teleportation is required, with a total e-bit cost of 3 with the same size constraints as in Figure \ref{fig:cig_so}. Edges which have been merged onto a control node are coloured red.}
        \label{fig:cig_eo}
\end{figure}
 
For condition 1., the shared control must be the same for each gate. Since there may be many controls involved in gate groups which are not the shared control, we refer to the qubit representing the shared control as the ``lead control qubit".The node representing the first appearance of the shared control is referred to as the ``original control node" for a group. ``Adjacent", in condition 2., means that there are no non-diagonal single-qubit gates acting on the original control between any consecutive gates in the group. From section \ref{sec:egt}, we know that diagonal single-qubit gates on the original control can form part of a group, so these can be freely applied to a qubit while it is linked to other QPUs. For symmetric two-qubit gates, in which both involved qubits are control qubits, the conditions are only relevant to the lead control qubit. There are no restrictions on the partner control qubits.

\begin{figure*}[ht!]
    \centering
    \begin{subfigure}[t]{0.5\textwidth}
    \centering
        \includegraphics[width=\textwidth]{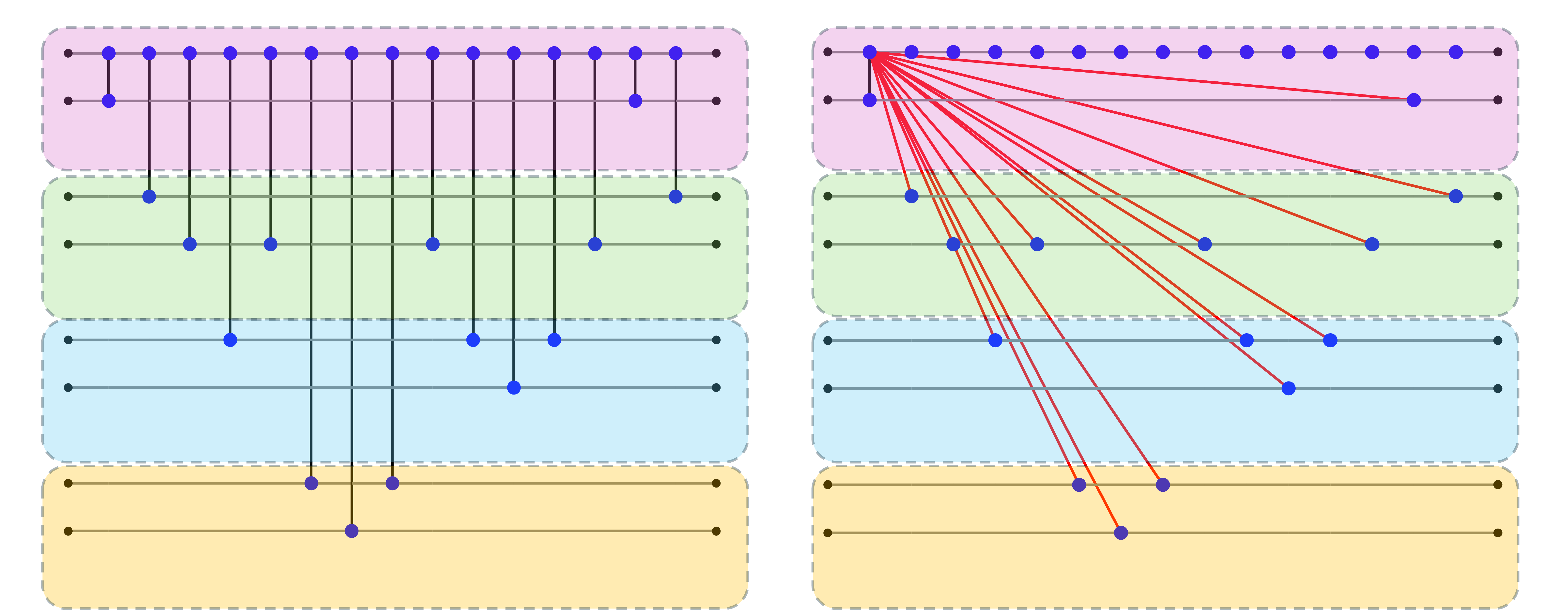}
        \caption{\textbf{Gate grouping.} All edges are merged onto the original control node. Since this node connects to 3 partitions outside its own, the e-bit cost for the assignment is 3.}
        \label{fig:gg}
    \end{subfigure}%
    ~
    \begin{subfigure}[t]{0.47\textwidth}
    \centering
        \includegraphics[width=\textwidth]{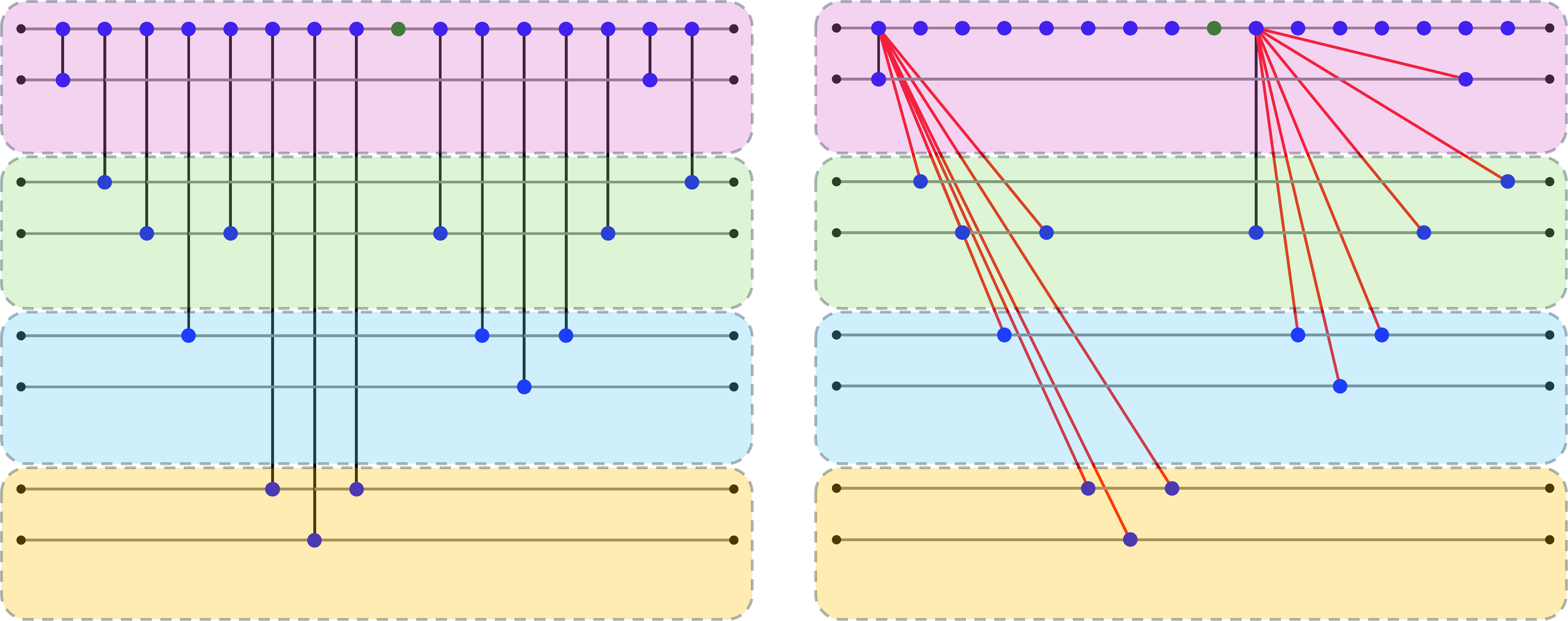}
        \caption{\textbf{Interrupted gate grouping.} Since the control line is interrupted by a non-diagonal single-qubit gate (green node), two gate groups must be formed. The resulting e-bit cost is 5, since the first group connects to 3 partitions and the second to 2.}
        \label{fig:gg_sep}
    \end{subfigure}
    
\end{figure*}

The edges for the gate groups identified are all merged onto the original control node. This is demonstrated in Figure \ref{fig:cig_eo}. Edges grouped on a particular original control node can be treated similarly to hyperedges in a hypergraph. This is because the e-bit requirements for distributing a group of gates will correspond not to the number of edges but rather the number of different QPUs to which the edges in the group connect to, since e-bits can be re-used within each linked QPU. This modification allows us to explore the full solution space from assignments which imply state teleportation only to those which rely on extended gate teleportation as well as everything in between. The new cost function is:

 \begin{equation} \label{cost_e}
c = \sum_{u,v \in V}e_{u,v}(1 - \delta_{\Phi(u)\Phi(v)}) h(u,\Phi(v))f(r(\Phi(u),\Phi(v))),
 \end{equation}
where $h(u,\Phi(v))$ is a function which returns 1 if $e_{u,v}$ is the first edge from source node $u$ connecting to a node in $\Phi(v)$. With this modification, we do not count edges which can benefit from an existing linked copy. To see why this is the case, consider the natural extension of the gate teleportation procedure in Section \ref{sec:state_telep}. In order for a qubit to be linked to $k$ QPUs, its state must be transformed to to a $(k+1)$-fold entangled state $\ket{\psi} = \alpha \ket{0}^{\otimes k+1} + \beta \ket{1}^{\otimes k+1}$, such that each linked communication qubit can be used to control operations for the lead control. Creating such a state requires the gate teleportation starting procedure to be repeated for each QPU, requiring $k$ e-bits to be consumed. Thus, for each group of edges which have been merged onto a control node, the e-bit cost is equal to the number of external QPUs which are connected. Once the links are active, the states can either remain at the communication qubit, or swap onto idle data qubits to free up communication qubit space. An example of grouping gates and calculating costs is shown in Figures \ref{fig:gg} and \ref{fig:gg_sep}.

 \section{Solution methods} 
 \subsection{Genetic algorithm}
 To solve GCP, we propose a genetic algorithm (GA). GAs are evolutionary algorithms which are well known for being able to produce good heuristic solutions in combinatorial optimisation problems \cite{muhlenbein_evolution_1988}, particularly if tailored by prior knowledge. In this case, we can constrain solutions of the problems to take the shape of matrices, where the column index corresponds to qubit number $j$, the row index to the layer $i$ of the circuit. Corresponding elements are assignments $\Phi(v)$ where $v$ corresponds to qubit $q_{j}$ at layer $i$. Each solution corresponds to a set of state and gate teleportations, shown in \ref{fig:assignment}. We can explore all valid solutions with permutations of the elements within each layer.

\begin{algorithm}
\caption{Genetic Circuit Partitioning}\label{alg:example}
\begin{algorithmic}[1]
\State \textbf{Input:} A graph $G(V,E)$ for circuit of depth $d$ with $n_{q}$ qubits. A set of QPUs $Q$ and their sizes $|Q_{n}|$. Population size $|L|$. Number of generations/iterations $N_{g}$.
\State \textbf{Output:} Sorted population $\tilde{L}$ of candidate assignments.
\State Initialise empty population $L$
\For{$n = 0$ \textbf{to} $|L| - 1$}
    \State Produce a single-layer, static, assignment $\phi$.
    \State Repeat $\phi$ for depth $d$ to create full assignment $\Phi$.
    \State Add candidate to $L$
\EndFor
\For{$n = 1$ \textbf{to} $N_{g} - 1$}
    \For{candidate \textbf{in} $L$}
        \State cost $\gets costfunc(G,\text{candidate})$
        \State Add cost to $costlist$
    \EndFor
    \State Initialise empty population $\tilde{L}$
    \For{$m = 0$ \textbf{to} $\frac{|L|}{2} - 1$}
        \State $dist \gets softmax(costlist)$
        \State Sample $c_a$ and $c_b$ from $L$ according to $dist$.
        \State $\tilde{c_a}, \tilde{c_b} \gets crossover(c_a, c_b)$
        \State $\tilde{c_a}, \tilde{c_b} \gets mutation(\tilde{c_a}, \tilde{c_b})$
        \State Add $\tilde{c_a}$, $\tilde{c_b}$ to $\tilde{L}$
    \EndFor
    \State $L \gets \tilde{L}$
\EndFor
\State \Return $\tilde{L}$
\end{algorithmic}
\end{algorithm}

The workflow for GCP is described in algorithm \ref{alg:example}. For $costfunc$, the cost functions in equations \ref{cost_s} and \ref{cost_e} are used, depending on which variant of the problem is being considered.The $crossover$ and $mutation$ functions can be chosen to suit the problem. We use a single point crossover, which selects crossover points by row of the matrix, so that the size constraints remain satisfied. For mutation, we use a custom mutation function which samples random pairs of nodes in a given interval and exchanges them if they have a net difference of neighbours assigned to each other's partition, inspired by the Kernighan-Lin heuristic \cite{kernighan_efficient_1970} This mutation function is slower than a normal flip mutation but better avoids local minima.

\subsection{Complexity analysis} \label{sec:complexity}

The time complexity of the genetic algorithm is determined by population size $|L|$, number of generations $N_{g}$, cost function evaluation and the crossover and mutation functions. For all tests, we used constant $L$ and $N_{g}$ such that the scaling is independent of these factors. Typically, as the size of the problem increases, a constant population size and number of iterations is less likely to reach to an optimal solution, however, we retain competitive and often superior performance to the benchmark methods with constant parameters in all tests. In this case the time complexity is $\mathcal{O}(cost + selection + crossover + mutation)$. The cost function requires checking the assignment of nodes for $n_{q}(d+1)$ state edges and a maximum of $\frac{1}{2}n_{q}d$ gate edges, such that the cost function time complexity is $\mathcal{O}(n_{q}d)$. The selection employs a softmax over all fitness values which has complexity $\mathcal{O}(|L|)$ which is constant since $|L|$ is fixed. Crossover simply selects a random point at which to split two solutions which is also a constant time operation. In our case, the mutation function checks the partitions of a subset of nodes for two qubits in a randomly chosen interval. This is done for a constant $k$ pairs of qubits, such that we query the partition of at worst $2dk$ nodes, making the complexity $\mathcal{O}(d)$. Altogether this gives a time complexity $\mathcal{O}(n_{q}d) + \mathcal{O}(d) = \mathcal{O}(n_{q}d)$, such that the scaling is linear in the overall circuit size. The complexity for specific circuits is discussed in section \ref{sec:disc}.


\section{Results}

\begin{figure*}[t!]
    \centering
    \begin{subfigure}[t]{0.24\textwidth}
        \centering
        \includegraphics[width=\textwidth]{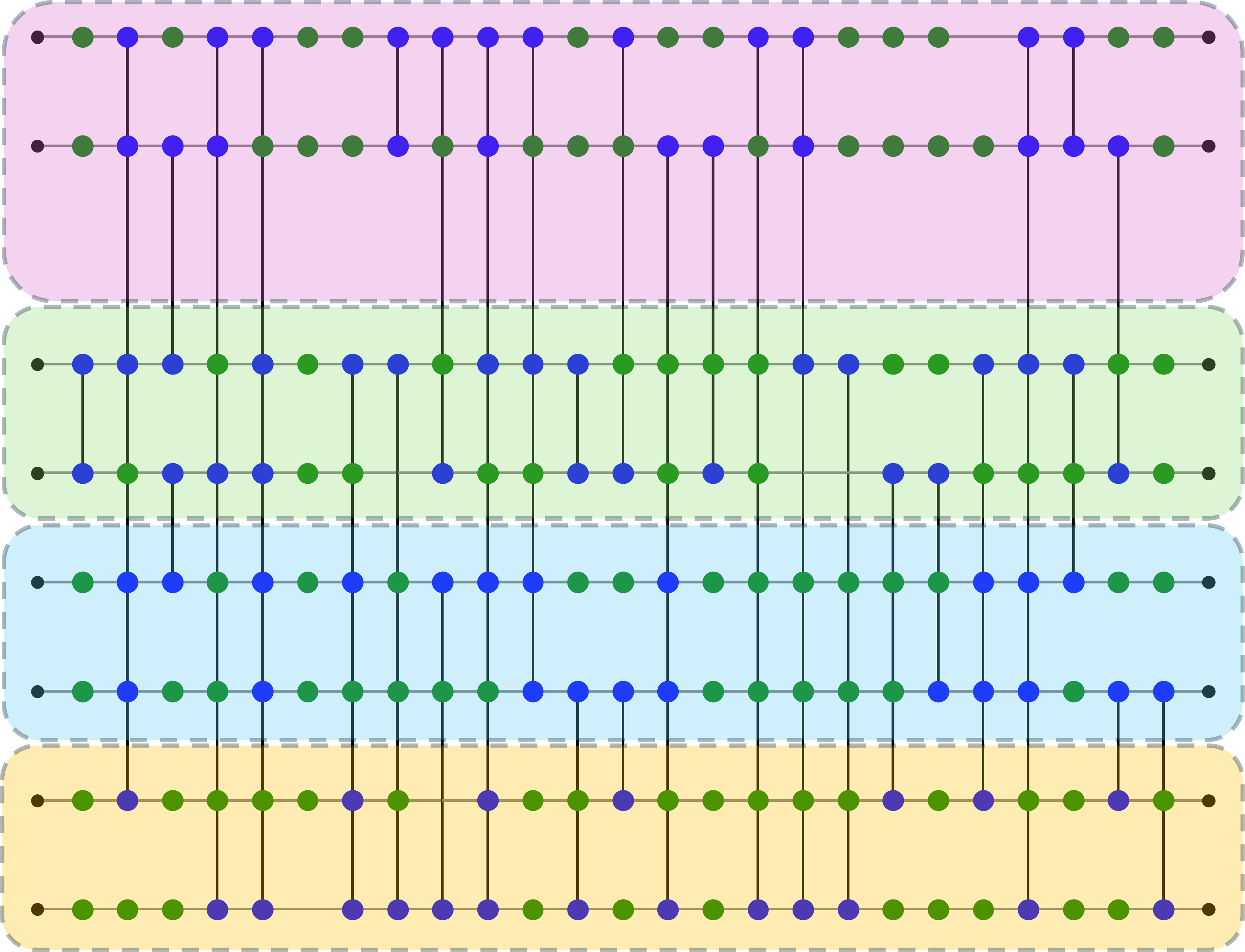}
        \caption{Unoptimised mapping. E-bit cost of 41.}
    \end{subfigure}%
    ~ 
    \begin{subfigure}[t]{0.24\textwidth}
        \centering
        \includegraphics[width=\textwidth]{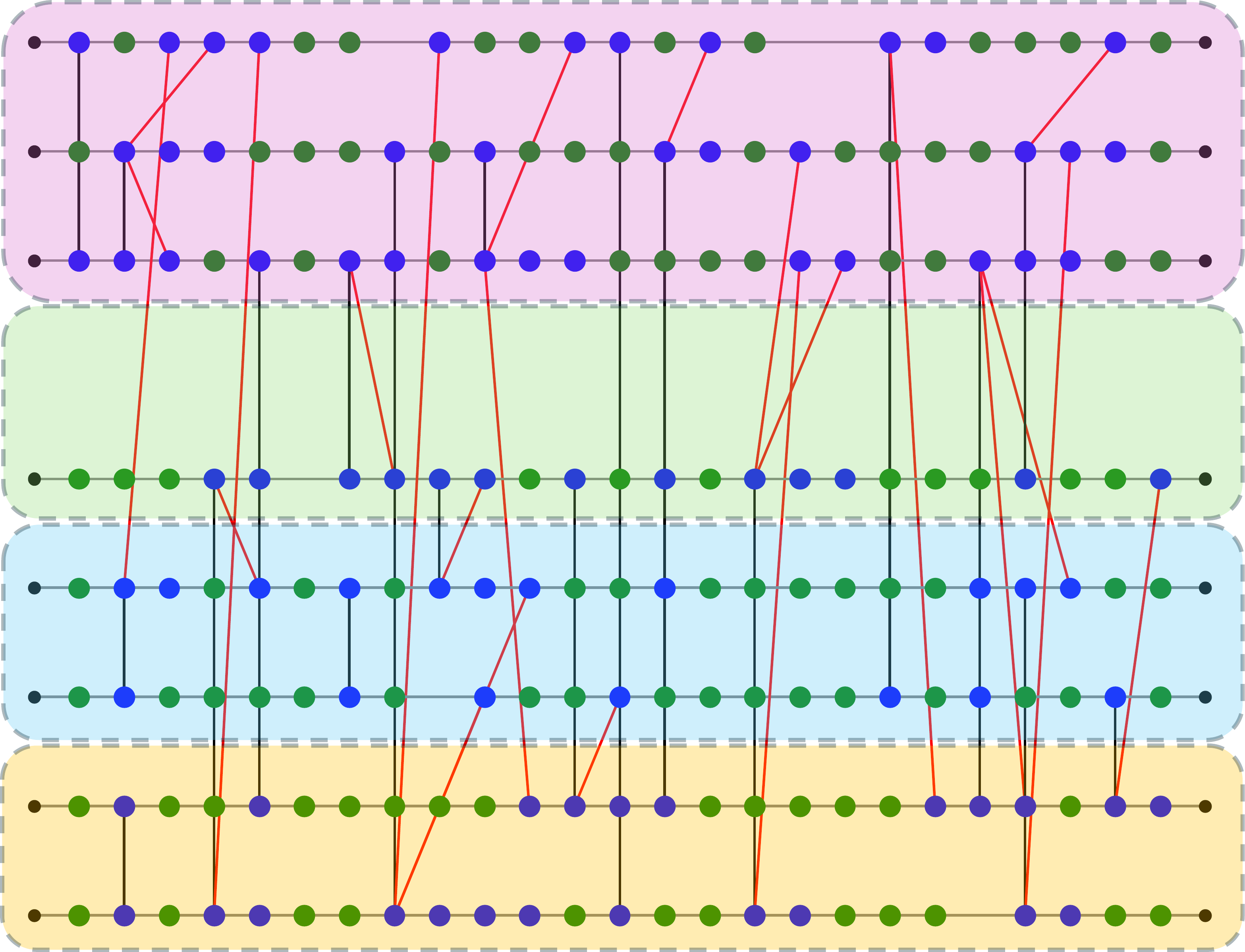}
        \caption{Static mapping using extended gate teleportation. E-bit cost of 28.}
    \end{subfigure}
    \begin{subfigure}[t]{0.24\textwidth}
        \centering
        \includegraphics[width=\textwidth]{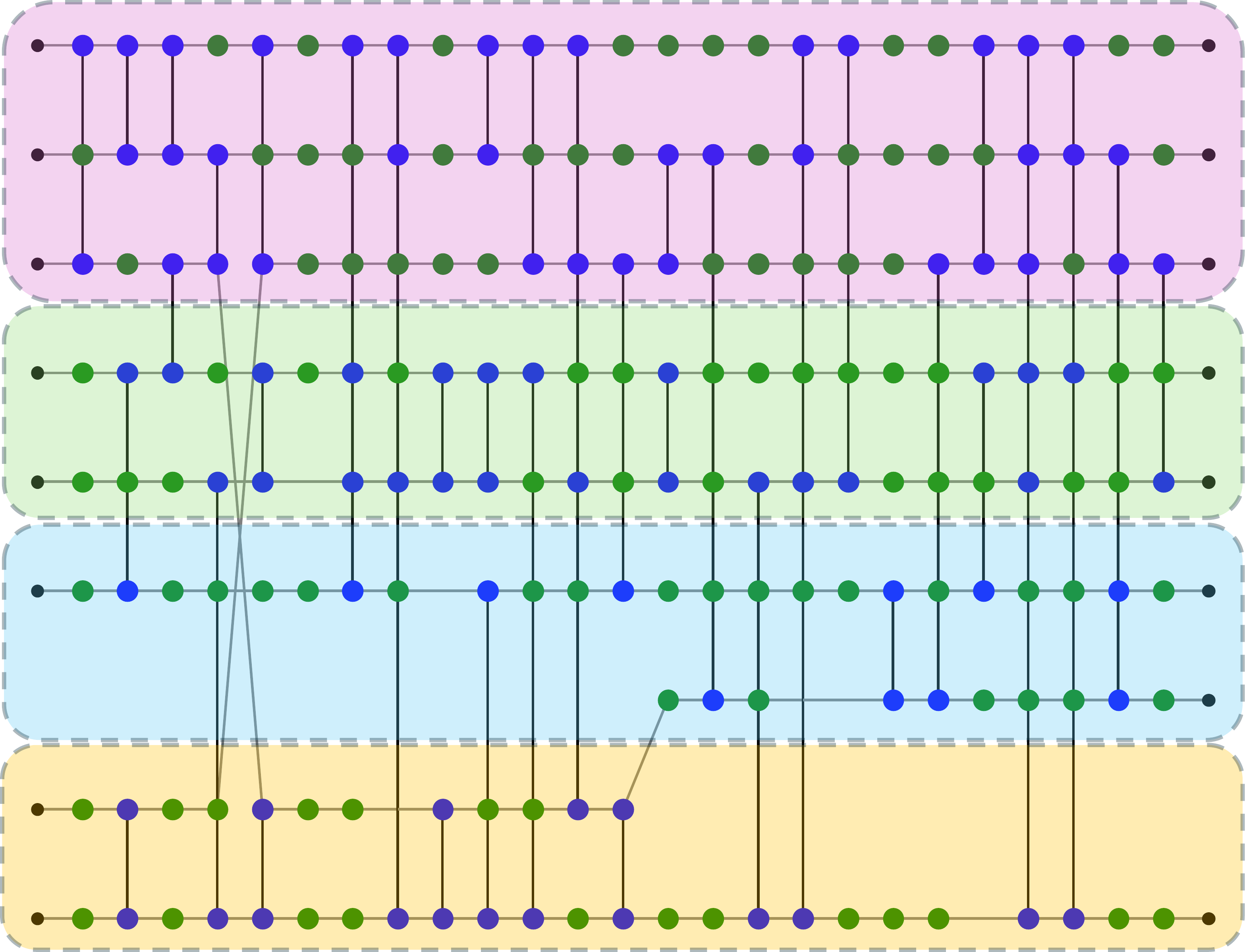}
        \caption{Sequential mapping using state and gate teleportation. E-bit cost of 29.}
    \end{subfigure}%
    \label{fig:mapping}
    \begin{subfigure}[t]{0.24\textwidth}
        \centering
        \includegraphics[width=\textwidth]{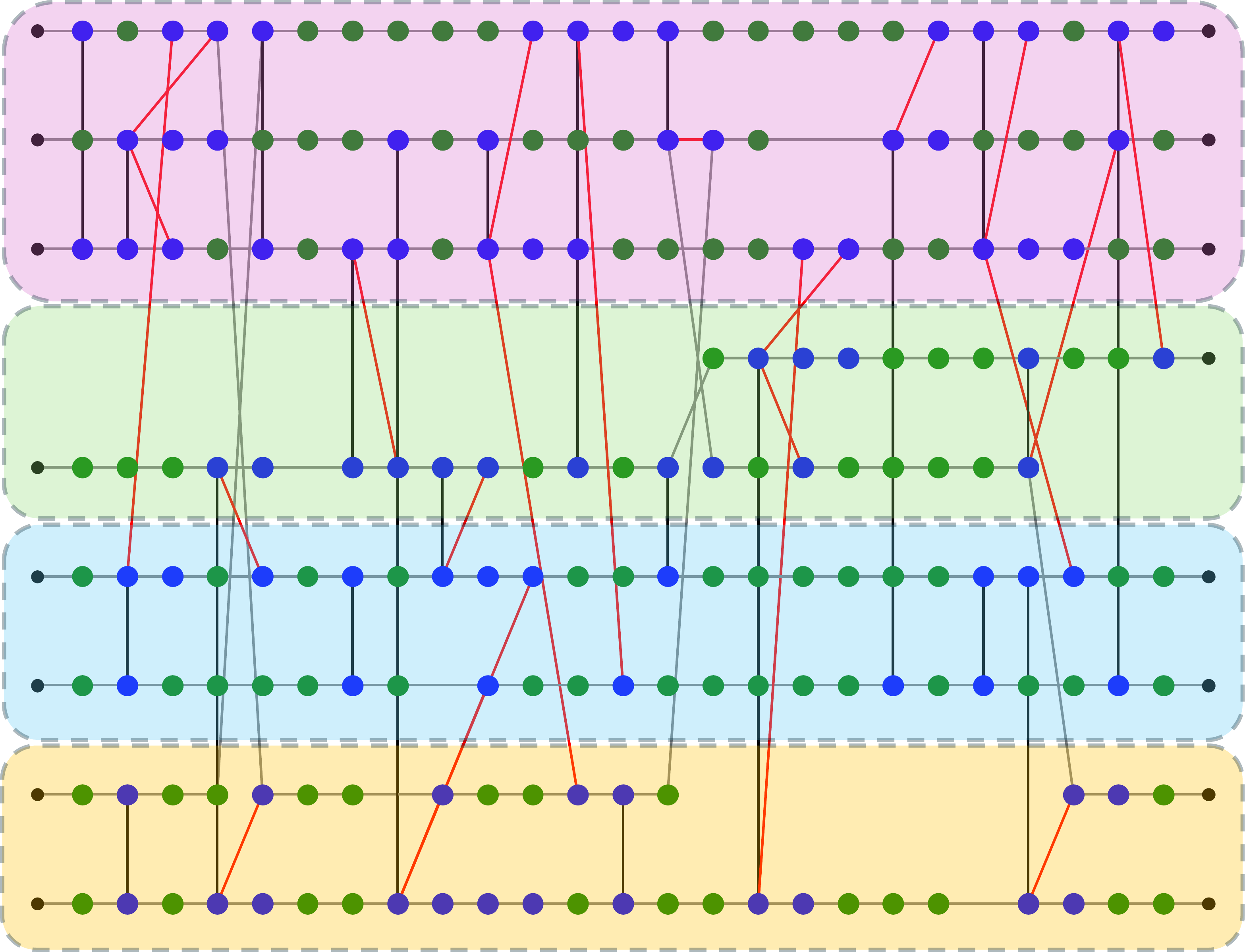}
        \caption{Sequential mapping using state teleportation and extended gate teleportation. E-bit cost of 25.}
    \end{subfigure}%
    \caption{Example of a CP fraction circuit mapped using different techniques. The best e-bit cost is in d) where all teleportation procedures are optimised over.}
\end{figure*}

We use the GA to produce results for both the simple and the extended variants of GCP, here termed GCP-S and GCP-E respectively. These are compared against two leading methods in the literature, exemplifying the extremes of state and gate teleportation-focused algorithms. The comparison graphs are shown in Figures \ref{fig:real_res}, \ref{fig:cp_frac_res} and \ref{fig:large}, while the results are summarised in Table \ref{tab:table}. All results are produced using an HP Elite Tower 800 G9, running Windows 11 Pro with an Intel Core i5-12500 processor, and 16GB of DDR5 RAM.
\subsection{Benchmark algorithms}
\subsubsection{\textbf{FGP-rOEE}}
The first is the algorithm presented in Baker et al. for solving their FGP problem formulation. The algorithm is a variant of the overall extreme exchange (OEE) graph partitioning heuristic \cite{park_algorithms_1995}, called relaxed-OEE. We use a version of rOEE which differs only in the initial phase, where the authors use the regular OEE algorithm to find a starting assignment. We use a simplified version of this which simply swaps the assignment of nodes until no benefit can be gained, then proceed with rOEE.

\subsubsection{\textbf{Pytket-DQC}}
The second algorithm comes from Andres-Martinez et al. \cite{andres-martinez_distributing_2023} and is freely available in pytket-DQC \cite{andres-martinez_cqclpytket-dqc_2024}. The authors present a number of different workflows \textendash ~we use a custom workflow which is the same as the best performing ``EmbedSteinerDetached (ESD)" but that it uses the annealing allocator rather than the hypergraph partitioner. We will refer to this workflow as AESD. Using the annealer may result in a small performance decrease, however, reliance on external software led to implementation issues.
\subsection{Benchmark circuits}
Four different circuit types are used for comparing methods, each with a very different qubit interaction structure. The choices are motivated such that there are instances where each of the benchmark algorithms performs both strongly and poorly. For all circuits, the universal gate set of controlled-phase gates $CP(\vartheta)$ and general unitary rotations $U(\phi,\theta,\lambda)$, is used. This gate set is useful for DQC as the $CP(\vartheta)$ allows for many gate packing possibilities. In all cases, circuits are partitioned among QPUs containing 8 qubits each.
\subsubsection{\textbf{CP fraction}}\label{sec:cp_frac}
CP fraction circuits are a generalisation of the \textit{CZ fraction} introduced in Sundaram et al. \cite{sundaram_efficient_2021} and used in later works \cite{sundaram_distribution_2022, andres-martinez_distributing_2023}. A $CZ$ fraction circuit, for a given number of qubits $n_{q}$, depth $d$ and fraction $p$, consists of $d$ layers of gates where, for each layer, a Hadamard ($H$) gate is applied to each qubit with probability $1 - p$, while all qubits for which $H$ was not applied are randomly paired with $CZ$ gates. The only difference in a $CP$ fraction is that each $H$ is replaced with $U(\phi,\theta,\lambda)$ for randomly chosen $\phi, \theta$ and $\lambda$. In addition, each $CZ$ is replaced with $CP(\vartheta)$, for some randomly chosen $\vartheta$. In all tests, we set $d = n_{q}$.

\subsubsection{\textbf{QFT}}
The QFT is a well-known subroutine in many quantum algorithms \cite{shor_algorithms_1994}, implemented using a very regular structure of two-qubit gates\cite{coppersmith_approximate_2002}. The QFT has been used as a benchmark for various circuit partitioning methods and is an example of a circuit type consisting of large distributable packets.

\begin{figure*}[ht!]
    \centering
    \begin{subfigure}[t]{0.3\textwidth}
        \centering
        \includegraphics[width=\textwidth]{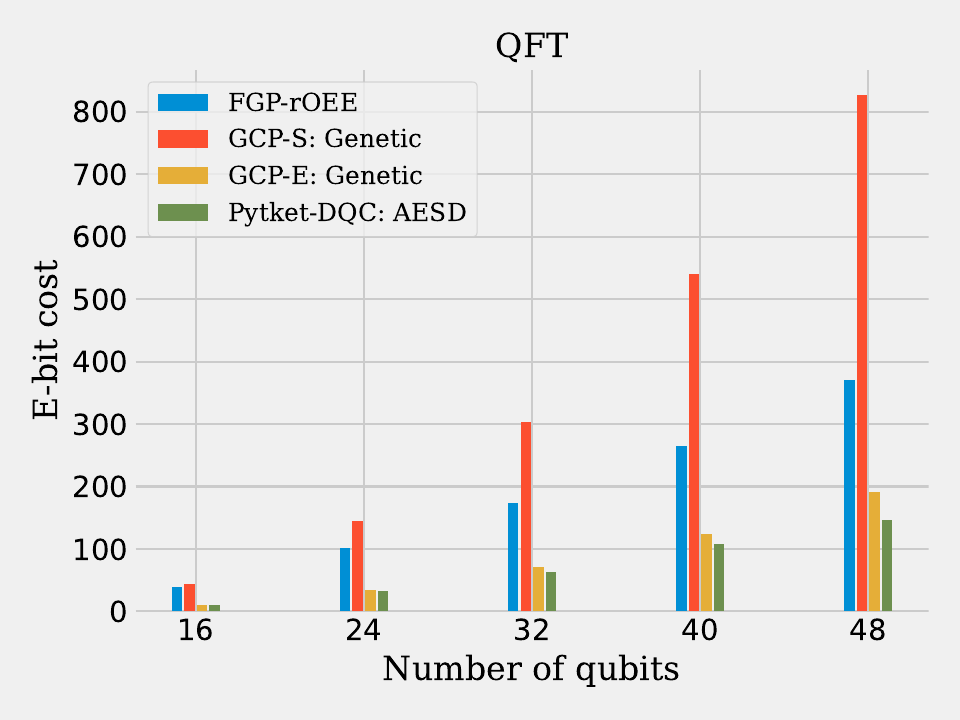}
    \end{subfigure}%
    ~ 
    \begin{subfigure}[t]{0.3\textwidth}
        \centering
        \includegraphics[width=\textwidth]{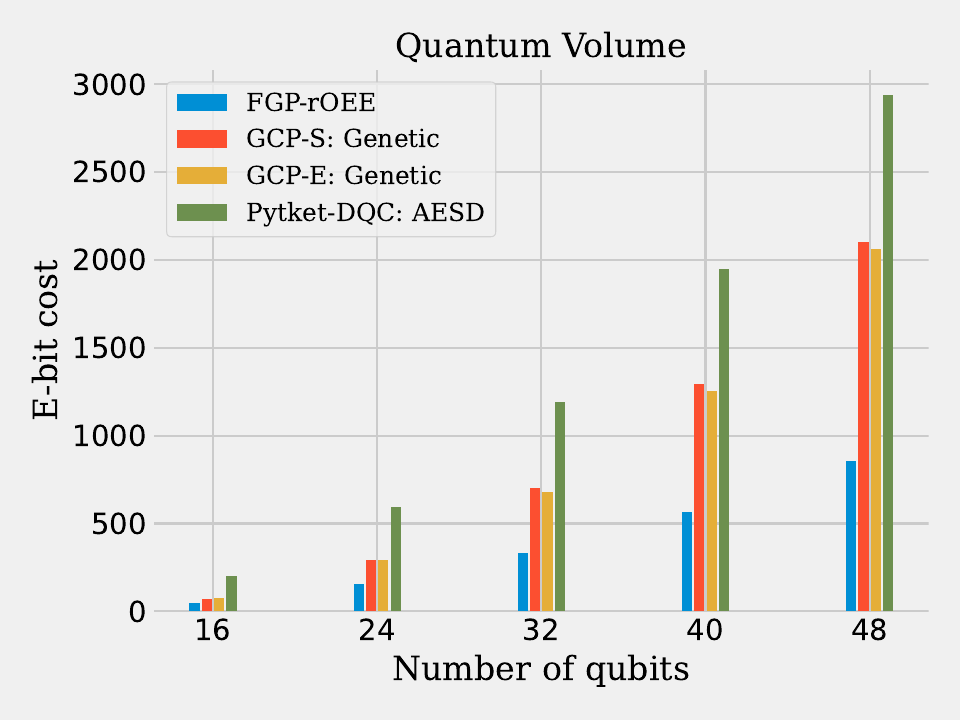}
        \caption{E-bit cost results}
    \end{subfigure}%
    ~ 
    \begin{subfigure}[t]{0.3\textwidth}
        \centering
        \includegraphics[width=\textwidth]{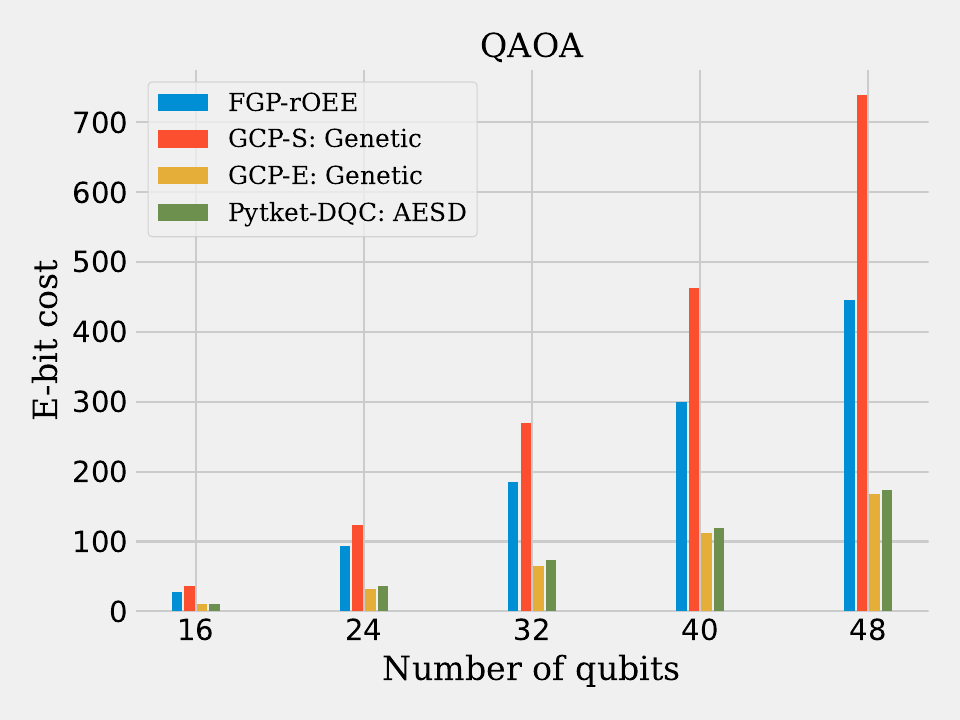}
    \end{subfigure}%
    
    \label{fig:e-bit_R}
    
    \begin{subfigure}[t]{0.3\textwidth}
        \centering
        \includegraphics[width=\textwidth]{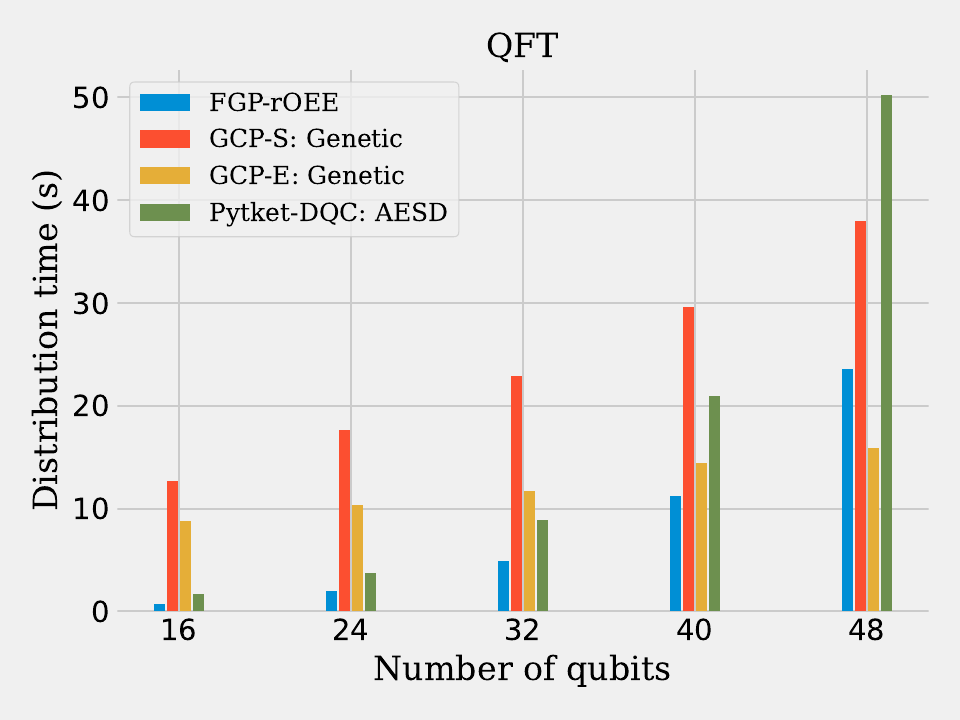}
    \end{subfigure}
    ~ 
    \begin{subfigure}[t]{0.3\textwidth}
        \centering
        \includegraphics[width=\textwidth]{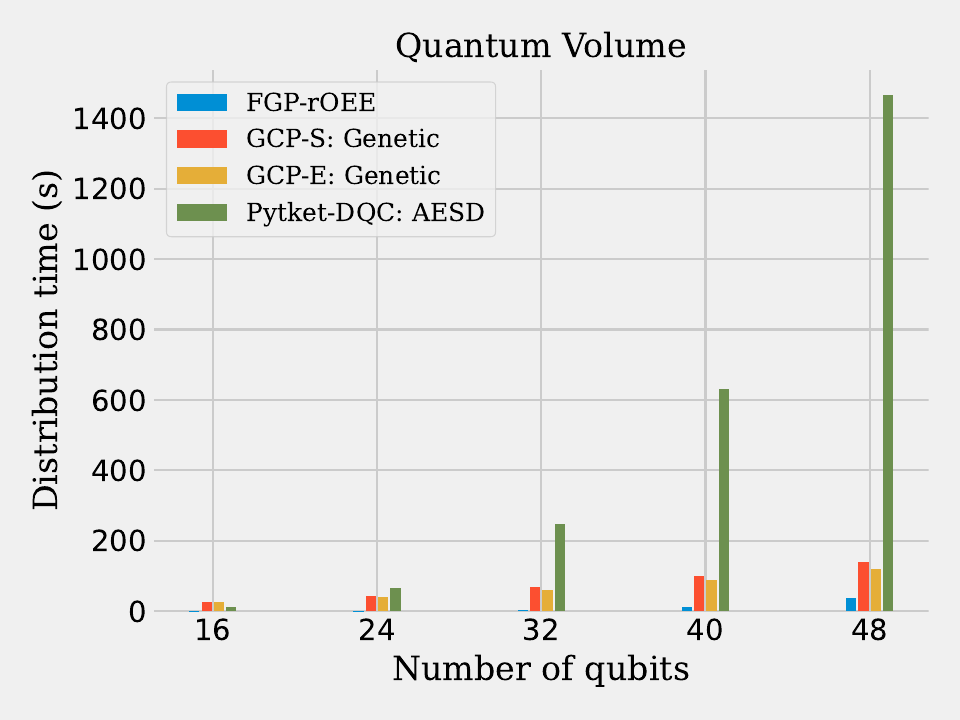}
        \caption{Time taken to generate distribution}
    \end{subfigure}
    ~ 
    \begin{subfigure}[t]{0.3\textwidth}
        \centering
        \includegraphics[width=\textwidth]{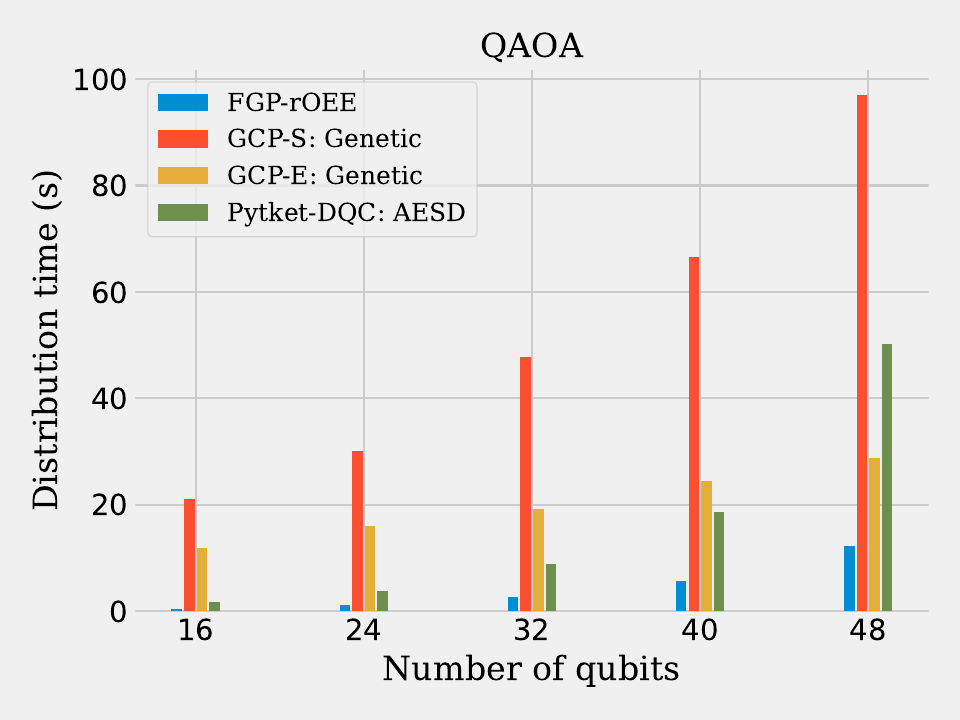}
    \end{subfigure}
    \label{fig:time_R}
\caption{Results from real circuits distributed over QPUs consisting of 8 qubits each.}
\label{fig:real_res}
\end{figure*}

\subsubsection{\textbf{Quantum volume (QV)}}
QV circuits are randomised circuits used for benchmarking the performance of quantum computers \cite{cross_validating_2019}. It is reasonable to believe that QV circuits are also a strong indicator of performance of a distributed quantum architecture, where good compiling strongly impacts the QV score. Another reason for choosing QV circuits for benchmarking is that they represent a middle ground between highly structured and fully random circuits. QV circuits consist of sequences of Haar random unitaries acting on pairs of qubits interleaved with permutations of qubit numbers \cite{cross_validating_2019}. The resulting circuit consists of alternating blocks of repeated single- and two-qubit gate patterns.

\subsubsection{\textbf{Quantum approximate optimisation algorithm (QAOA)}}

QAOA circuits are designed to solve combinatorial optimisation problems using parameterised quantum circuits\cite{farhi_quantum_2014}. The circuit consists of parameterised single- and two-qubit gates which are classically optimised to find an approximate solution to the problem. The structure of the circuit depends partly on the structure of the problem being solved. We consider randomly initialised QAOA circuits for MaxCut problems over random graphs with an edge probability of 50\% for each pair of nodes.

\begin{figure*}[t!]
    \centering
    \begin{subfigure}[t]{0.3\textwidth}
        \centering
        \includegraphics[width=\textwidth]{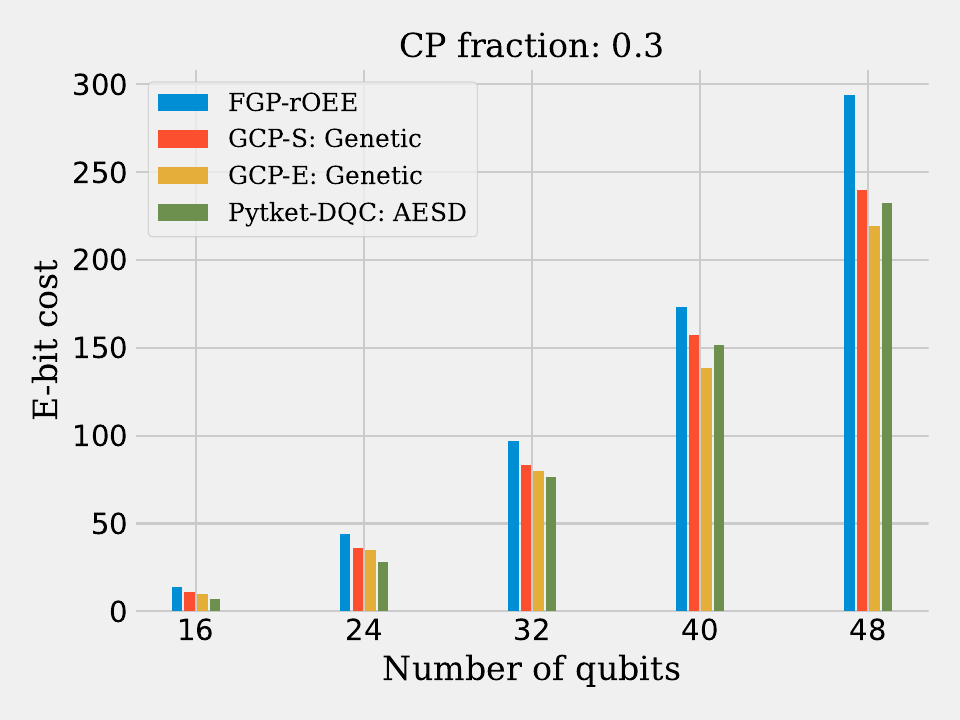}
    \end{subfigure}%
    ~ 
    \begin{subfigure}[t]{0.3\textwidth}
        \centering
        \includegraphics[width=\textwidth]{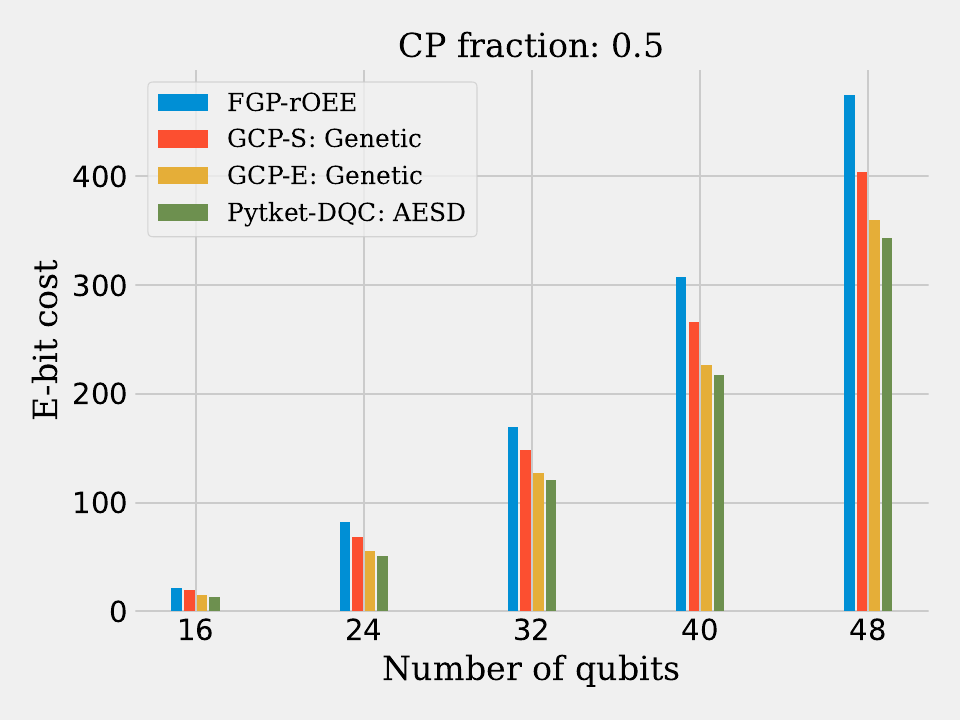}
    \end{subfigure}%
    ~ 
    \begin{subfigure}[t]{0.3\textwidth}
        \centering
        \includegraphics[width=\textwidth]{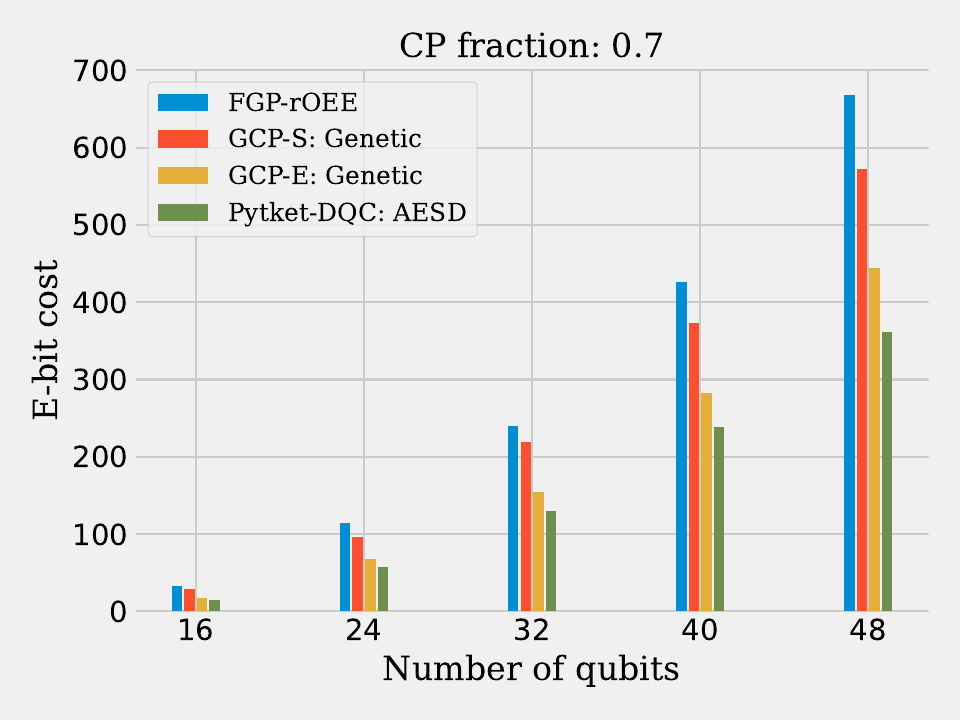}

    \end{subfigure}%
    \label{fig:e-bit_CP}
    \caption*{}
    ~
    \begin{subfigure}[t]{0.3\textwidth}
        \centering
        \includegraphics[width=\textwidth]{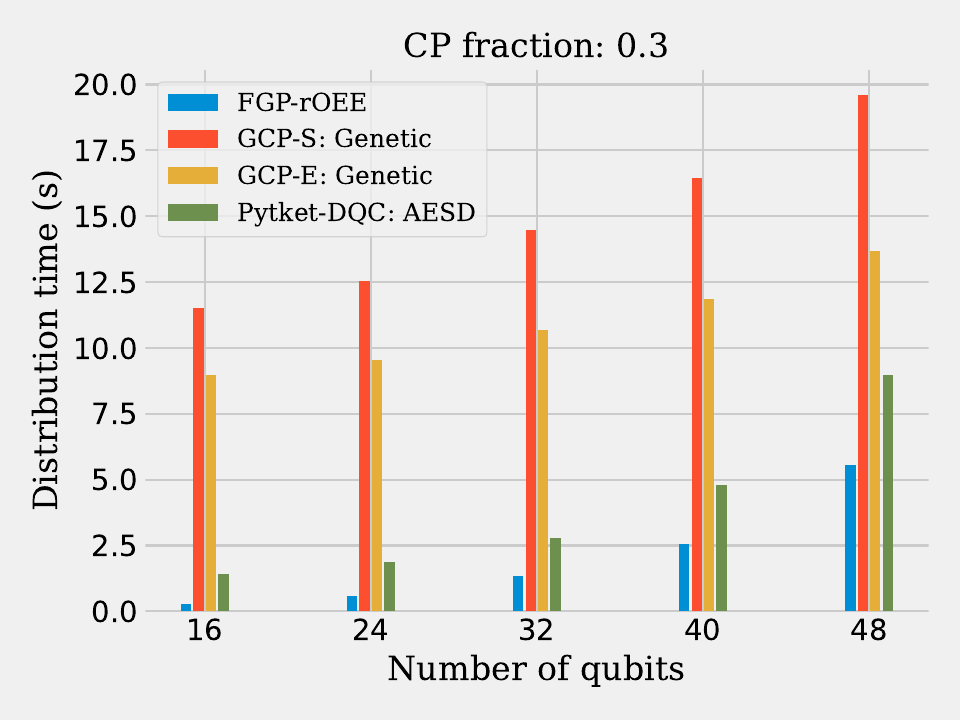}
    \end{subfigure}
    ~ 
    \begin{subfigure}[t]{0.3\textwidth}
        \centering
        \includegraphics[width=\textwidth]{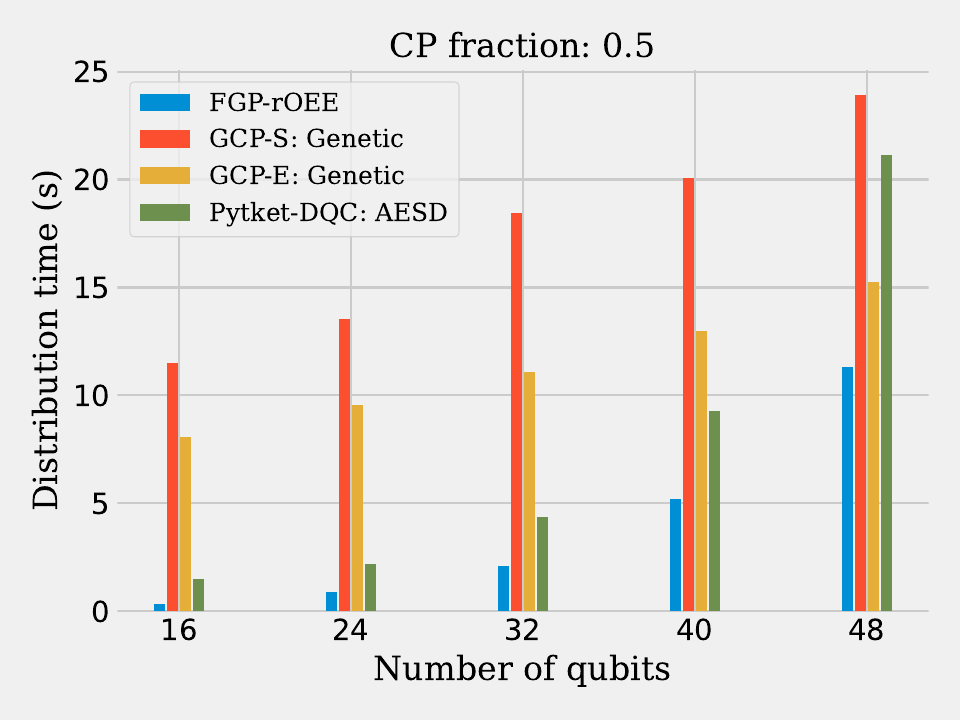}
    \end{subfigure}
    ~ 
    \begin{subfigure}[t]{0.3\textwidth}
        \centering
        \includegraphics[width=\textwidth]{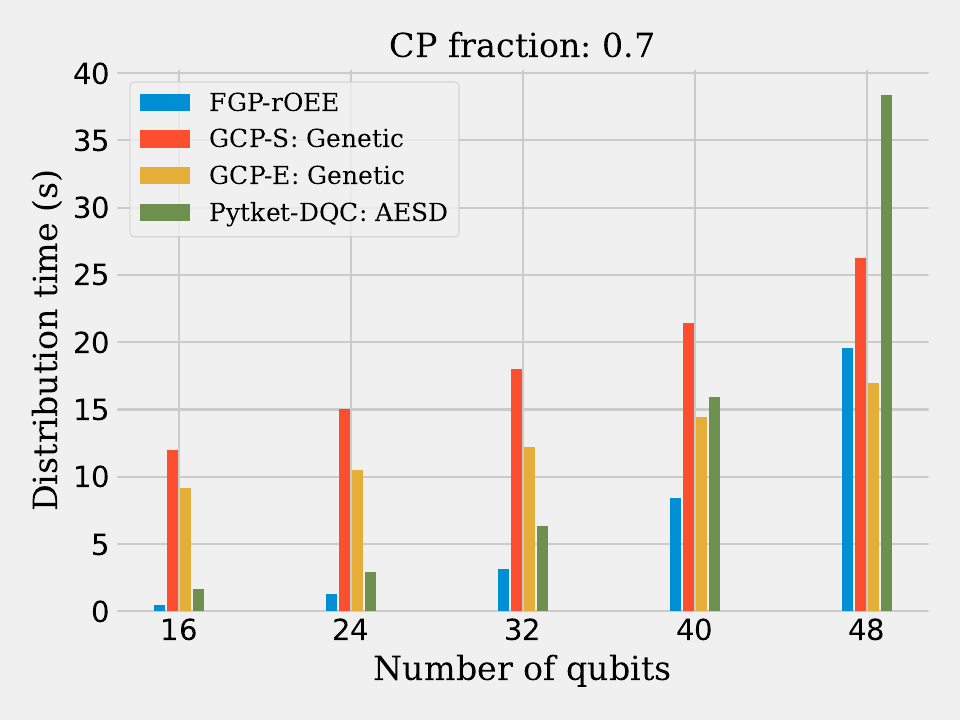}
        
    \end{subfigure}
    \label{fig:time_CP}
\caption{Results for random, CP fraction circuits, as described in section \ref{sec:cp_frac}, distributed over QPUs consisting of 8 qubits each}
\label{fig:cp_frac_res}
\end{figure*}

\begin{figure*}[t!]
    \centering
    \begin{subfigure}[t]{0.35\textwidth}
        \centering
        \includegraphics[width=\textwidth]{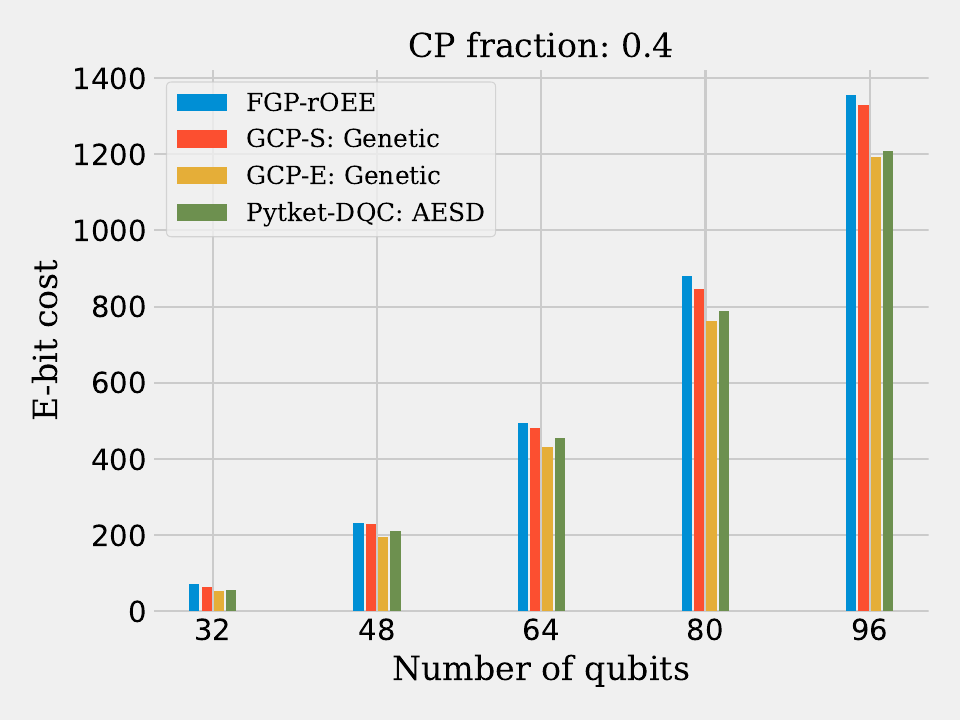}
    \end{subfigure}%
    ~ 
    \begin{subfigure}[t]{0.35\textwidth}
        \centering
        \includegraphics[width=\textwidth]{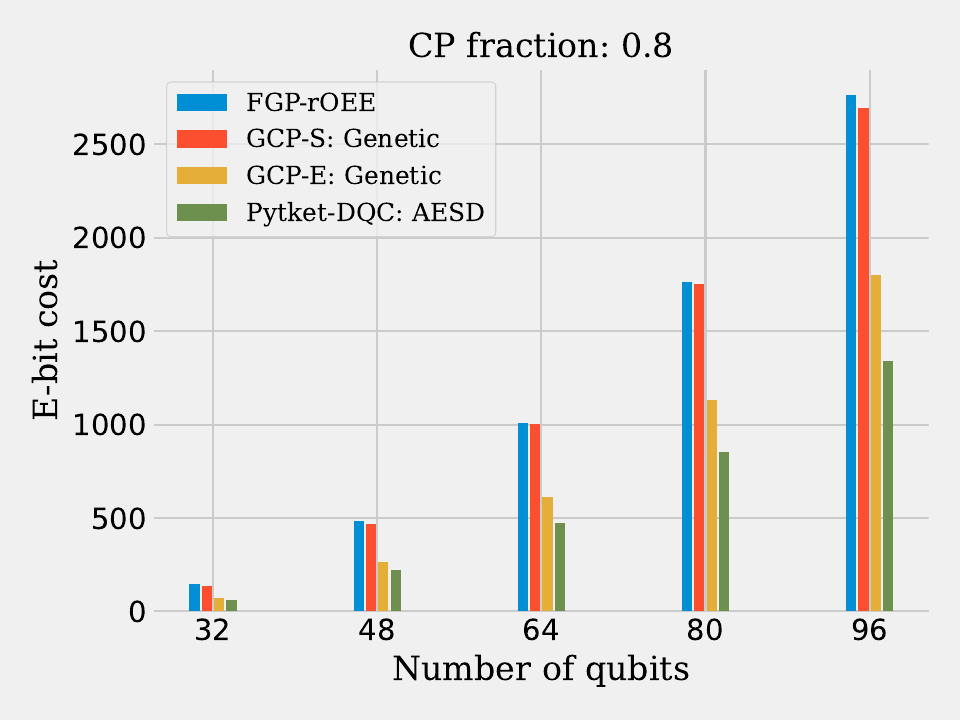}
    \end{subfigure}
    \caption*{}
    \begin{subfigure}[t]{0.35\textwidth}
        \centering
        \includegraphics[width=\textwidth]{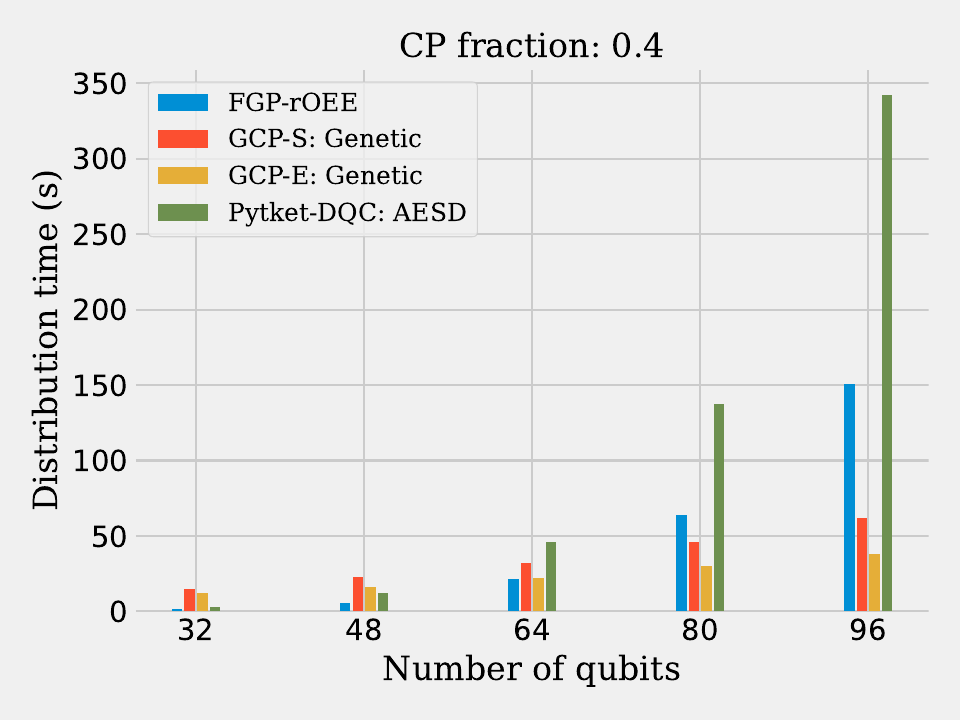}
    \end{subfigure}%
    ~
    \begin{subfigure}[t]{0.35\textwidth}
        \centering
        \includegraphics[width=\textwidth]{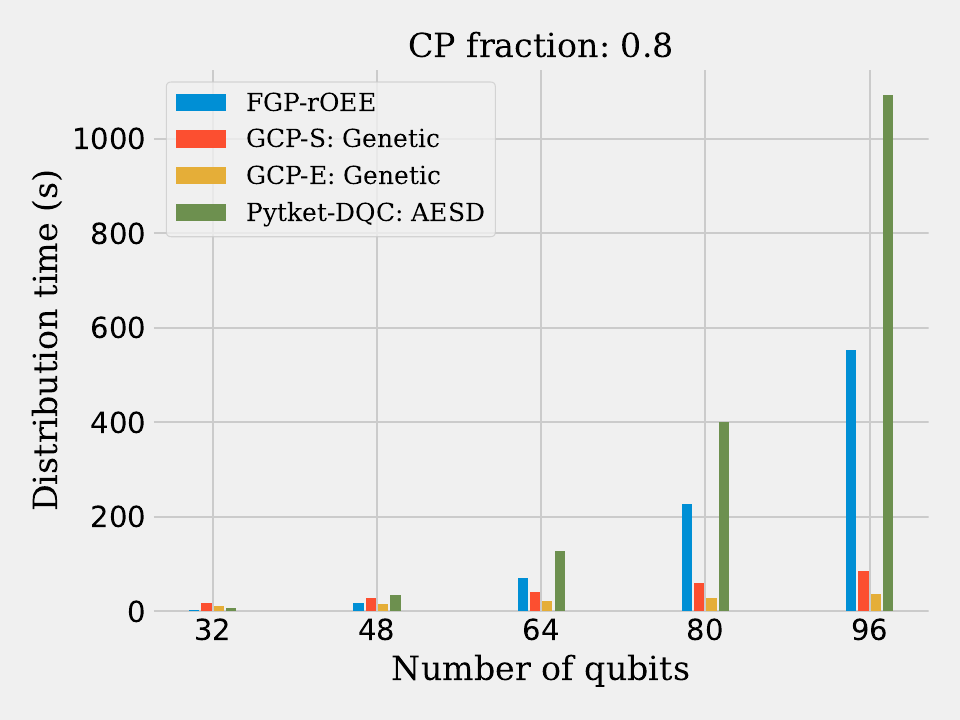}
    \end{subfigure}%
\caption{Results from larger CP fraction circuits distributed over QPUs consisting of 16 qubits each.}
\label{fig:large}
\end{figure*}

\begin{table}[H]
    \centering
    \begin{tabular}{lcccc}
        \toprule
        \textbf{Method} & \textbf{FGP-rOEE} & \textbf{GCP-S} & \textbf{GCP-E} & \textbf{AESD} \\
        \midrule
        CP 0.3 & 0.66 & 0.54 & 0.50 & 0.49 \\
        CO 0.4 & 0.59 & 0.57 & 0.50 & 0.52 \\
        CP 0.5 & 0.65 & 0.55 & 0.47 & 0.45 \\
        CP 0.7 & 0.66 & 0.57 & 0.42 & 0.35 \\
        CP 0.8	& 0.59 & 0.58 & 0.36 & 0.28 \\
        QFT & 0.34 & 0.59  & 0.14 & 0.12 \\
        QV & 0.20 & 0.43 & 0.42 & 0.73 \\
        QAOA & 0.35 & 0.51 & 0.12 & 0.14 \\
        \midrule
        \textbf{Average} & 0.52 & 0.55 & 0.37 & 0.39 \\

        \bottomrule
        
    \end{tabular}
    \caption{Average performance across all tests. The metric used is the ratio of e-bits required to the total number of two-qubit gates in the circuit, such that a lower score indicates a more efficient circuit distribution.}
    \label{tab:table}
\end{table}
\section{Discussion and conclusion}\label{sec:disc}
It is clear from the results in Figures \ref{fig:real_res}, \ref{fig:cp_frac_res} and \ref{fig:large} that methods which depend heavily on state teleportation or gate teleportation perform very well in specific circumstances but do not retain the performance in all cases. For example, AESD performs very well on QFT circuits, as these have multiple long chains of two-qubit gates which share a common control qubit. As a result, no state teleportation is required for efficient distribution. However, the AESD performs poorly on QV circuits, which do not contain any distributable packets. In this case, these methods are limited to providing a qubit assignment which performs only as well as static graph partitioning. In contrast, FGP-rOEE performs exceptionally well on QV circuits, where full reliance on state teleportation is always most effective. This is because all blocks of a QV circuit can be made local if state teleportations are inserted to match the qubit permutation between each section. This makes it particularly easy for FGP-rOEE to find an optimal sequence of assignments. In principle, both GCP-S and GCP-E should be able to find the same assignment, but it becomes increasingly less likely as the size of the circuit increases so would require longer runtime for the GA. Again, since there are no distributable packets in QV circuits, GCP-E reduces to GCP-S which is reflected in the similarity of their performance. 

For $CP$ fraction circuits, which have a more random structure, FGP-rOEE performs the worst, particularly as the number of qubits and fraction of two-qubit gates increases. Since these circuits have non-uniform qubit interactions, it can be inefficient to be constrained to state teleportation. In contrast, AESD performs worse than GCP-E when the fraction of two-qubit gates is lower. This is because there are more single-qubit gates separating the two-qubit gates, which decreases the average size of the distributable packets. GCP-E can benefit from state teleportation in this cases so retains low e-bit cost. As the fraction of two-qubit gates increases however, AESD performs the best in terms of e-bit cost. AESD contains the most flexible conditions for extended gate teleportation, with the benefits of embedding reflected for two-qubit gate fractions greater than 0.4. This improved performance comes at a significant time cost, however, which can be seen in the distribution time graphs. For all circuits containing 40 qubits or more, the distribution time of AESD drastically increases. In these instances, the cost performance of GCP-E remains competitive for a much shorter distribution time. The overall, average performance is shown in Table \ref{tab:table}. GCP-E is the best performer in terms of e-bit cost, outperforming FGP-rOEE by around 20\% and AESD by around 2\%. 

The time graphs in Figures \ref{fig:real_res}, \ref{fig:cp_frac_res} and \ref{fig:large} accord with the scaling from section \ref{sec:complexity}. While slower at smaller scales, GCP-S and GCP-E scale better and are the fastest in most cases by the end. CP fractions circuits are built with $d = n_{q}$, while QFT and QV circuits have $d \propto n_{q}$, so in these cases the complexity reduces to $\mathcal{O}(n_{q}^{2})$. For QAOA, the depth is more variable and depends on the optimisation problem. Since we use circuits designed for MaxCut on random graphs, the depth scales with the number of edges of the graphs and the number of repetitions, so steeper polynomial scaling is expected. The experimental results demonstrate superior scaling compared to the benchmark methods for most circuits. However, the complexities of these methods are highly dependent on the type of circuit. The results from larger circuits, in Figure \ref{fig:large}, show significant time improvements over the benchmark methods.

In future, we look to extend these methods to general network topologies. We avoided doing this here, as current methods for doing so (scaling cost linearly with path distance \cite{sundaram_distributing_2023,andres-martinez_distributing_2023}) are over-simplified and require careful treatment. We acknowledge another assumption regarding the number of communication qubits available to facilitate e-bit generation at each QPU. At present, this assumption can be justified on the basis that communication qubits can freely swap states in and out of data qubits which are idle in computation. For realistic systems with data and communication qubit number constraints, restrictions will need to be enforced and used to guide the optimisation process.

\section{Acknowledgements}
The authors thank Richard Meister for engaging discussions and acknowledge funding from the Engineering and Physical Sciences Research Council (EPSRC) funded Distributed Quantum Computing project, grant number EP/W032643/1

\bibliographystyle{siamurl}
\bibliography{filtered_references}

\begin{thebibliography}{10}

\bibitem{cuomo_towards_2020}
{\sc D.~Cuomo, M.~Caleffi and A.~S. Cacciapuoti}, {\em Towards a distributed quantum computing ecosystem}, IET Quantum Communication {\bfseries 1} (2020), pp.~3--8.
\newblock \_eprint: https://onlinelibrary.wiley.com/doi/pdf/10.1049/iet-qtc.2020.0002.
\newblock {\sc url: }\url{https://onlinelibrary.wiley.com/doi/abs/10.1049/iet-qtc.2020.0002}, \href {http://dx.doi.org/10.1049/iet-qtc.2020.0002} {\path{doi:10.1049/iet-qtc.2020.0002}}.

\bibitem{caleffi_distributed_2022}
{\sc M.~Caleffi et~al.}, {\em Distributed {Quantum} {Computing}: a {Survey}}, Dec. 2022.
\newblock arXiv:2212.10609 [quant-ph].
\newblock {\sc url: }\url{http://arxiv.org/abs/2212.10609}, \href {http://dx.doi.org/10.48550/arXiv.2212.10609} {\path{doi:10.48550/arXiv.2212.10609}}.

\bibitem{unruh_maintaining_1995}
{\sc W.~G. Unruh}, {\em Maintaining coherence in quantum computers}, Physical Review A {\bfseries 51} (1995), pp.~992--997.
\newblock Publisher: American Physical Society.
\newblock {\sc url: }\url{https://link.aps.org/doi/10.1103/PhysRevA.51.992}, \href {http://dx.doi.org/10.1103/PhysRevA.51.992} {\path{doi:10.1103/PhysRevA.51.992}}.

\bibitem{eisert_optimal_2000}
{\sc J.~Eisert, K.~Jacobs, P.~Papadopoulos and M.~B. Plenio}, {\em Optimal local implementation of nonlocal quantum gates}, Physical Review A {\bfseries 62} (2000), p.~052317.
\newblock Publisher: American Physical Society.
\newblock {\sc url: }\url{https://link.aps.org/doi/10.1103/PhysRevA.62.052317}, \href {http://dx.doi.org/10.1103/PhysRevA.62.052317} {\path{doi:10.1103/PhysRevA.62.052317}}.

\bibitem{isailovic_interconnection_2006}
{\sc N.~Isailovic, Y.~Patel, M.~Whitney and J.~Kubiatowicz}, {\em Interconnection {Networks} for {Scalable} {Quantum} {Computers}}, ACM SIGARCH Computer Architecture News {\bfseries 34} (2006), pp.~366--377.
\newblock {\sc url: }\url{https://dl.acm.org/doi/10.1145/1150019.1136505}, \href {http://dx.doi.org/10.1145/1150019.1136505} {\path{doi:10.1145/1150019.1136505}}.

\bibitem{andres-martinez_distributing_2023}
{\sc P.~Andres-Martinez et~al.}, {\em Distributing circuits over heterogeneous, modular quantum computing network architectures}, July 2023.
\newblock arXiv:2305.14148 [quant-ph].
\newblock {\sc url: }\url{http://arxiv.org/abs/2305.14148}.

\bibitem{barral_review_2024}
{\sc D.~Barral et~al.}, {\em Review of {Distributed} {Quantum} {Computing}. {From} single {QPU} to {High} {Performance} {Quantum} {Computing}}, 2024.
\newblock Publisher: arXiv Version Number: 1.
\newblock {\sc url: }\url{https://arxiv.org/abs/2404.01265}, \href {http://dx.doi.org/10.48550/ARXIV.2404.01265} {\path{doi:10.48550/ARXIV.2404.01265}}.

\bibitem{baker_time-sliced_2020}
{\sc J.~M. Baker, C.~Duckering, A.~Hoover and F.~T. Chong}, {\em Time-{Sliced} {Quantum} {Circuit} {Partitioning} for {Modular} {Architectures}}, May 2020.
\newblock {\sc url: }\url{https://arxiv.org/abs/2005.12259v1}, \href {http://dx.doi.org/10.1145/3387902.3392617} {\path{doi:10.1145/3387902.3392617}}.

\bibitem{sundaram_distributing_2023}
{\sc R.~G. Sundaram and H.~Gupta}, {\em Distributing {Quantum} {Circuits} {Using} {Teleportations}}, May 2023.
\newblock arXiv:2306.00195 [quant-ph].
\newblock {\sc url: }\url{http://arxiv.org/abs/2306.00195}, \href {http://dx.doi.org/10.48550/arXiv.2306.00195} {\path{doi:10.48550/arXiv.2306.00195}}.

\bibitem{nikahd_automated_2021}
{\sc E.~Nikahd, N.~Mohammadzadeh, M.~Sedighi and M.~S. Zamani}, {\em Automated window-based partitioning of quantum circuits}, Physica Scripta {\bfseries 96} (2021), p.~035102.
\newblock Publisher: IOP Publishing.
\newblock {\sc url: }\url{https://dx.doi.org/10.1088/1402-4896/abd57c}, \href {http://dx.doi.org/10.1088/1402-4896/abd57c} {\path{doi:10.1088/1402-4896/abd57c}}.

\bibitem{andres-martinez_automated_2019}
{\sc P.~Andrés-Martínez and C.~Heunen}, {\em Automated distribution of quantum circuits via hypergraph partitioning}, Physical Review A {\bfseries 100} (2019), p.~032308.
\newblock Publisher: American Physical Society.
\newblock {\sc url: }\url{https://link.aps.org/doi/10.1103/PhysRevA.100.032308}, \href {http://dx.doi.org/10.1103/PhysRevA.100.032308} {\path{doi:10.1103/PhysRevA.100.032308}}.

\bibitem{wu_entanglement-efficient_2023}
{\sc J.-Y. Wu et~al.}, {\em Entanglement-efficient bipartite-distributed quantum computing}, Quantum {\bfseries 7} (2023), p.~1196.
\newblock Publisher: Verein zur Förderung des Open Access Publizierens in den Quantenwissenschaften.
\newblock {\sc url: }\url{https://quantum-journal.org/papers/q-2023-12-05-1196/}, \href {http://dx.doi.org/10.22331/q-2023-12-05-1196} {\path{doi:10.22331/q-2023-12-05-1196}}.

\bibitem{cuomo_optimized_2023}
{\sc D.~Cuomo et~al.}, {\em Optimized {Compiler} for {Distributed} {Quantum} {Computing}}, ACM Transactions on Quantum Computing {\bfseries 4} (2023), pp.~1--29.
\newblock {\sc url: }\url{https://dl.acm.org/doi/10.1145/3579367}, \href {http://dx.doi.org/10.1145/3579367} {\path{doi:10.1145/3579367}}.

\bibitem{sundaram_efficient_2021}
{\sc R.~G. Sundaram}, {\em Efficient {Distribution} of {Quantum} {Circuits}}, 2021.

\bibitem{ferrari_compiler_2021}
{\sc D.~Ferrari, A.~S. Cacciapuoti, M.~Amoretti and M.~Caleffi}, {\em Compiler {Design} for {Distributed} {Quantum} {Computing}}, IEEE Transactions on Quantum Engineering {\bfseries 2} (2021), pp.~1--20.
\newblock {\sc url: }\url{https://ieeexplore.ieee.org/document/9334411/}, \href {http://dx.doi.org/10.1109/TQE.2021.3053921} {\path{doi:10.1109/TQE.2021.3053921}}.

\bibitem{ferrari_modular_2023}
{\sc D.~Ferrari, S.~Carretta and M.~Amoretti}, {\em A {Modular} {Quantum} {Compilation} {Framework} for {Distributed} {Quantum} {Computing}}, IEEE Transactions on Quantum Engineering {\bfseries 4} (2023), pp.~1--13.
\newblock arXiv:2305.02969 [quant-ph].
\newblock {\sc url: }\url{http://arxiv.org/abs/2305.02969}, \href {http://dx.doi.org/10.1109/TQE.2023.3303935} {\path{doi:10.1109/TQE.2023.3303935}}.

\bibitem{karypis_fast_1998}
{\sc G.~Karypis and V.~Kumar}, {\em A {Fast} and {High} {Quality} {Multilevel} {Scheme} for {Partitioning} {Irregular} {Graphs}}, SIAM Journal on Scientific Computing {\bfseries 20} (1998), pp.~359--392.
\newblock {\sc url: }\url{http://epubs.siam.org/doi/10.1137/S1064827595287997}, \href {http://dx.doi.org/10.1137/S1064827595287997} {\path{doi:10.1137/S1064827595287997}}.

\bibitem{bennett_teleporting_1993}
{\sc C.~H. Bennett, G.~Brassard, C.~Crépeau, R.~Jozsa, A.~Peres and W.~K. Wootters}, {\em Teleporting an unknown quantum state via dual classical and {Einstein}-{Podolsky}-{Rosen} channels}, Physical Review Letters {\bfseries 70} (1993), pp.~1895--1899.
\newblock Publisher: American Physical Society.
\newblock {\sc url: }\url{https://link.aps.org/doi/10.1103/PhysRevLett.70.1895}, \href {http://dx.doi.org/10.1103/PhysRevLett.70.1895} {\path{doi:10.1103/PhysRevLett.70.1895}}.

\bibitem{huelga_quantum_2001}
{\sc S.~F. Huelga, J.~A. Vaccaro, A.~Chefles and M.~B. Plenio}, {\em Quantum remote control: {Teleportation} of unitary operations}, Physical Review A {\bfseries 63} (2001), p.~042303.
\newblock Publisher: American Physical Society.
\newblock {\sc url: }\url{https://link.aps.org/doi/10.1103/PhysRevA.63.042303}, \href {http://dx.doi.org/10.1103/PhysRevA.63.042303} {\path{doi:10.1103/PhysRevA.63.042303}}.

\bibitem{yimsiriwattana_generalized_2004}
{\sc A.~Yimsiriwattana and S.~J. Lomonaco~Jr}, {\em Generalized {GHZ} {States} and {Distributed} {Quantum} {Computing}}, Mar. 2004.
\newblock arXiv:quant-ph/0402148.
\newblock {\sc url: }\url{http://arxiv.org/abs/quant-ph/0402148}.

\bibitem{davis_towards_2023}
{\sc M.~G. Davis, J.~Chung, D.~Englund and R.~Kettimuthu}, {\em Towards {Distributed} {Quantum} {Computing} by {Qubit} and {Gate} {Graph} {Partitioning} {Techniques}}, Oct. 2023.
\newblock arXiv:2310.03942 [quant-ph].
\newblock {\sc url: }\url{http://arxiv.org/abs/2310.03942}.

\bibitem{muhlenbein_evolution_1988}
{\sc H.~Mühlenbein, M.~Gorges-Schleuter and O.~Krämer}, {\em Evolution algorithms in combinatorial optimization}, Parallel Computing {\bfseries 7} (1988), pp.~65--85.
\newblock {\sc url: }\url{https://www.sciencedirect.com/science/article/pii/0167819188900981}, \href {http://dx.doi.org/10.1016/0167-8191(88)90098-1} {\path{doi:10.1016/0167-8191(88)90098-1}}.

\bibitem{kernighan_efficient_1970}
{\sc B.~W. Kernighan and S.~Lin}, {\em An efficient heuristic procedure for partitioning graphs}, The Bell System Technical Journal {\bfseries 49} (1970), pp.~291--307.
\newblock Conference Name: The Bell System Technical Journal.
\newblock {\sc url: }\url{https://ieeexplore.ieee.org/document/6771089}, \href {http://dx.doi.org/10.1002/j.1538-7305.1970.tb01770.x} {\path{doi:10.1002/j.1538-7305.1970.tb01770.x}}.

\bibitem{park_algorithms_1995}
{\sc T.~Park and C.~Y. Lee}, {\em Algorithms for partitioning a graph}, Computers \& Industrial Engineering {\bfseries 28} (1995), pp.~899--909.
\newblock {\sc url: }\url{https://www.sciencedirect.com/science/article/pii/036083529500003J}, \href {http://dx.doi.org/10.1016/0360-8352(95)00003-J} {\path{doi:10.1016/0360-8352(95)00003-J}}.

\bibitem{andres-martinez_cqclpytket-dqc_2024}
{\sc P.~Andres-Martinez, D.~Mills, T.~Forrer and L.~Henaut}, {\em {CQCL}/pytket-dqc}, June 2024.
\newblock original-date: 2021-12-20T18:54:45Z.
\newblock {\sc url: }\url{https://github.com/CQCL/pytket-dqc}.

\bibitem{sundaram_distribution_2022}
{\sc R.~G. Sundaram, H.~Gupta and C.~R. Ramakrishnan}, {\em Distribution of {Quantum} {Circuits} {Over} {General} {Quantum} {Networks}}, June 2022.
\newblock arXiv:2206.06437 [quant-ph].
\newblock {\sc url: }\url{http://arxiv.org/abs/2206.06437}, \href {http://dx.doi.org/10.48550/arXiv.2206.06437} {\path{doi:10.48550/arXiv.2206.06437}}.

\bibitem{shor_algorithms_1994}
{\sc P.~Shor}, {\em Algorithms for quantum computation: discrete logarithms and factoring}, in Proceedings 35th {Annual} {Symposium} on {Foundations} of {Computer} {Science}, Nov. 1994, pp.~124--134.
\newblock {\sc url: }\url{https://ieeexplore.ieee.org/document/365700/?arnumber=365700}, \href {http://dx.doi.org/10.1109/SFCS.1994.365700} {\path{doi:10.1109/SFCS.1994.365700}}.

\bibitem{coppersmith_approximate_2002}
{\sc D.~Coppersmith}, {\em An approximate {Fourier} transform useful in quantum factoring}, Jan. 2002.
\newblock arXiv:quant-ph/0201067.
\newblock {\sc url: }\url{http://arxiv.org/abs/quant-ph/0201067}, \href {http://dx.doi.org/10.48550/arXiv.quant-ph/0201067} {\path{doi:10.48550/arXiv.quant-ph/0201067}}.

\bibitem{cross_validating_2019}
{\sc A.~W. Cross, L.~S. Bishop, S.~Sheldon, P.~D. Nation and J.~M. Gambetta}, {\em Validating quantum computers using randomized model circuits}, Physical Review A {\bfseries 100} (2019), p.~032328.
\newblock arXiv:1811.12926 [quant-ph].
\newblock {\sc url: }\url{http://arxiv.org/abs/1811.12926}, \href {http://dx.doi.org/10.1103/PhysRevA.100.032328} {\path{doi:10.1103/PhysRevA.100.032328}}.

\bibitem{farhi_quantum_2014}
{\sc E.~Farhi, J.~Goldstone and S.~Gutmann}, {\em A {Quantum} {Approximate} {Optimization} {Algorithm}}, Nov. 2014.
\newblock arXiv:1411.4028 [quant-ph].
\newblock {\sc url: }\url{http://arxiv.org/abs/1411.4028}.

\end{thebibliography}

\end{document}